\documentclass[twocolumn,showpacs,preprintnumbers,prl,fleqn,floatfix]{revtex4}

\usepackage{graphicx}
\usepackage{dcolumn}
\usepackage{bm}
\usepackage{amsmath, amssymb}

\begin{document}

\title{Hyperfine coupling and spin polarization in the bulk of the topological insulator Bi$_2$Se$_3$}

\author{S. Mukhopadhyay$^{1}$, S. Kr\"{a}mer$^{1}$, H. Mayaffre$^{1}$, H.F.
Legg$^{1}$, M. Orlita$^{1}$, C. Berthier$^{1}$, M.
Horvati\'c$^{1}$, G. Martinez$^{1}$, M. Potemski$^{1}$, B.A.
Piot$^{1}$}

\affiliation{$^{1}$ Laboratoire National des Champs Magn\'etiques
Intenses (LNCMI), CNRS-UJF-UPS-INSA, F-38042 Grenoble, France}

\author{A. Materna$^{2}$, G. Strzelecka$^{2}$, A. Hruban$^{2}$}

\affiliation{$^{2}$ Institute of electronic Materials Technology,
ul. Wolczynska 133, 01-919 Warsaw, Poland}

\date{\today}

\begin{abstract}

Nuclear magnetic resonance (NMR) and transport measurements have
been performed at high magnetic fields and low temperatures in a
series of $n$-type Bi$_{2}$Se$_{3}$ crystals. In low density
samples, a complete spin polarization of the electronic system is
achieved, as observed from the saturation of the isotropic
component of the $^{209}$Bi NMR shift above a certain magnetic
field. The corresponding spin splitting, defined in the
phenomenological approach of a 3D electron gas with a large
(spin-orbit-induced) effective $g$-factor, scales as expected with
the Fermi energy independently determined by simultaneous
transport measurements. Both the effective electronic $g$-factor
and the ``contact'' hyperfine coupling constant are precisely
determined. The magnitude of this latter reveals a non negligible
$s$-character of the electronic wave function at the bottom of the
conduction band. Our results show that the bulk electronic spin
polarization can be directly probed via NMR and pave the way for
future NMR investigations of the electronic states in Bi-based
topological insulators.

\end{abstract}
 \maketitle

Bismuth selenide, Bi$_{2}$Se$_{3}$, known for years as a narrow
gap semiconductor, has recently appeared as one of the first
examples of ``3D topological
insulators''\cite{Zhangtheo2009,XiaDiraccone2009,ChenBi2Te32009}.
As such unique state of matter, it is characterized by the
coexistence of 2-dimensional conducting surface states with an
insulating bulk material. The charge carriers at the surface
behave as massless relativistic particles (Dirac fermions) with a
spin locked to their translational momentum. These so-called
``helical Dirac fermions'', which promise applications in the
field of spintronic \cite{MacDospintro2012} and quantum
computation \cite{ReadQC2012}, have recently raised a considerable
interest (see Ref.~\onlinecite{Hasan2010} for a review). As a
matter of fact, the existence of gapless states at the boundary of
the material is related to a well-defined change in the
\textit{bulk} band structure.  In Bi$_{2}$Se$_{3}$, this
originates from a parity inversion of the valence and conduction
band in the presence of a large spin-orbit coupling
\cite{Zhangtheo2009}.

In an effort to deepen our understanding of the spin properties of
topological insulators, a characterization of the coupling between
the charge carriers and the nuclei in the Bi$_{2}$Se$_{3}$ matrix
is of high importance. Indeed, nuclear spins can inherently couple
to the topologically protected electronic states and limit their
coherence time. On the other hand, this hyperfine coupling can be
efficiently exploited to probe the electronic system via NMR
techniques. In particular, an electronic system bearing non-zero
spin polarization acts as an effective local magnetic field which
modifies the nuclei resonance frequency. This so-called ``Knight
shift'' has previously been extensively studied to probe the
electronic spin polarisation \cite{Barrett1995} as well as the
spatial symmetry of the wave functions
\cite{Leloup1973,Chekhovich2013} in some semiconductor-based bulk
or low dimensional systems. A couple of recent works have
investigated the NMR properties of Bi$_{2}$Se$_{3}$ samples and
revealed signatures of the bulk electronic states
\cite{Young2012,Nisson2013}. These measurements were however
limited to a narrow magnetic field and/or carrier density range,
and did not systematically include a complete transport
characterization of the bulk Fermi surface. The question of
whether or not the electron spin degree of freedom can be reliably
and directly accessed through such measurements is therefore still
pending, and constitute the main motivation of this work.

In this work, we present a large body of NMR and transport
measurements showing that the spin degree of freedom of bulk
electronic states in Bi$_{2}$Se$_{3}$ can be probed via the
hyperfine interaction with the Bi nuclei. This is evidenced by the
observation of specific high magnetic fields features in the
$^{209}$Bi NMR shift in $n$-type Bi$_{2}$Se$_{3}$ single crystals.
In the lowest density samples, the complete spin polarization of
the conduction electrons can be achieved, as indicated by the
saturation of the isotropic component of the NMR shift above a
density-dependent magnetic field. This allows us to precisely
estimate the ``contact term'' of the hyperfine coupling, providing
information on the spatial distribution of the electronic wave
function, and to determine the amplitude and sign of an effective
electronic $g$-factor describing the spin splitting in
Bi$_{2}$Se$_{3}$. The ``contact term'' is of the same order of
magnitude as in the typical semiconductors with \textit{s}-like
electron wave function in the conduction band, in spite of the
\textbf{$p$}-type nature of electrons in Bi$_{2}$Se$_{3}$
anticipated in the theoretical models near the bulk band gap of
this topological insulator \cite{Zhangtheo2009,Cao2013}. A more
complex magnetic field behavior is observed for the anisotropic
component of the Knight shift usually related to \textit{p}-like
electrons. Our results pave the way for future unconventional
experiments such as resistively \cite{Tiemann2012,Piot2012} or
optically-detected NMR aiming at testing the spin-physics of the
surfaces states in Bi-based topological insulators.

The Bi$_{2}$Se$_{3}$ crystals studied in this work were grown by
the vertical Bridgman method \cite{Hruban2011}. Several batches
with different amount of vacancies, known to act as a $n$-type
dopant in Bi$_{2}$Se$_{3}$, were prepared to obtain $n$-type
samples with different carrier concentration. A complete
characterization of the samples has been obtained by using
temperature- and angular-dependent magneto transport experiments
\cite{SM.I.} and shows that bulk transport is dominant in our
systems. Our sample pool cover a Fermi energy range spanning
18-137 meV from the bottom of the conduction band, corresponding
to electron densities $n_{e}$ from $7.5\times 10^{17}$cm$^{-3}$ to
$1.6\times 10^{19}$cm$^{-3}$. NMR measurements were performed both
in superconducting magnets (6.5 - 17 T) and in a 20 MW resistive
magnet (up to 30 T) \cite{SM.II.A}.

\begin{figure}[!h]
\begin{center}
\includegraphics[width= 7.5cm]{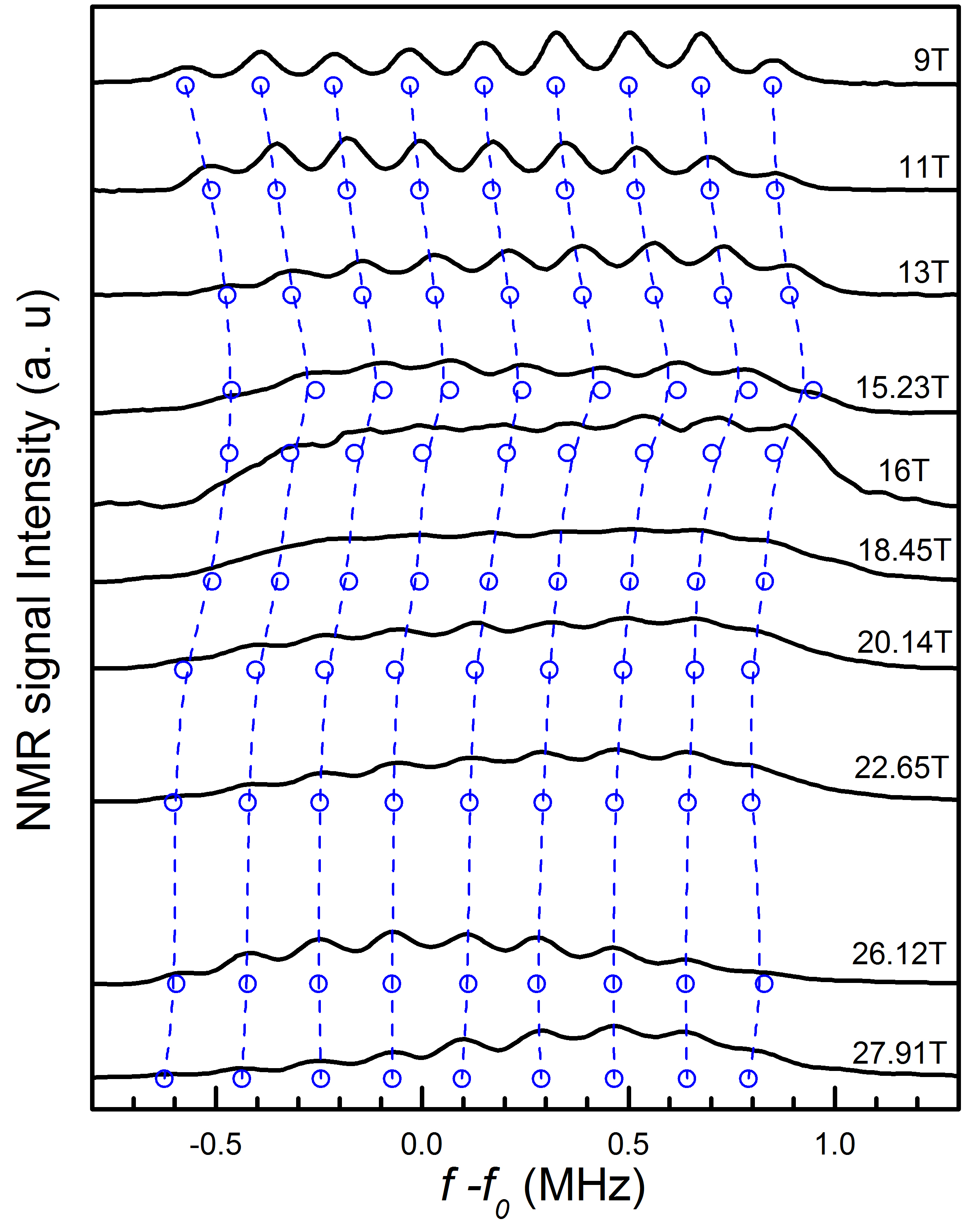}
\end{center}
\caption{(color online) Raw frequency-swept NMR spectra of sample
P3-6 ($n_{e}=7.5\times 10^{17}$cm$^{-3}$) as a function of the
magnetic field $B\parallel$ $c$-axis at $T=5$ K. The vertical
offset between the spectra is linear in $B$. The open circles
denote the fitted NMR line positions, while the connecting dash
lines are guide to eye. The frequency axis origin is defined by
f$_{0}= ^{209}\gamma B$, the resonance for the bare $^{209}$Bi
nuclei ($^{209}\gamma = 6.84184$ MHz/T
\cite{Carter1977}).}\label{Fig1}
\end{figure}
\begin{figure}[!h]
\begin{center}
\includegraphics[width=8.5cm]{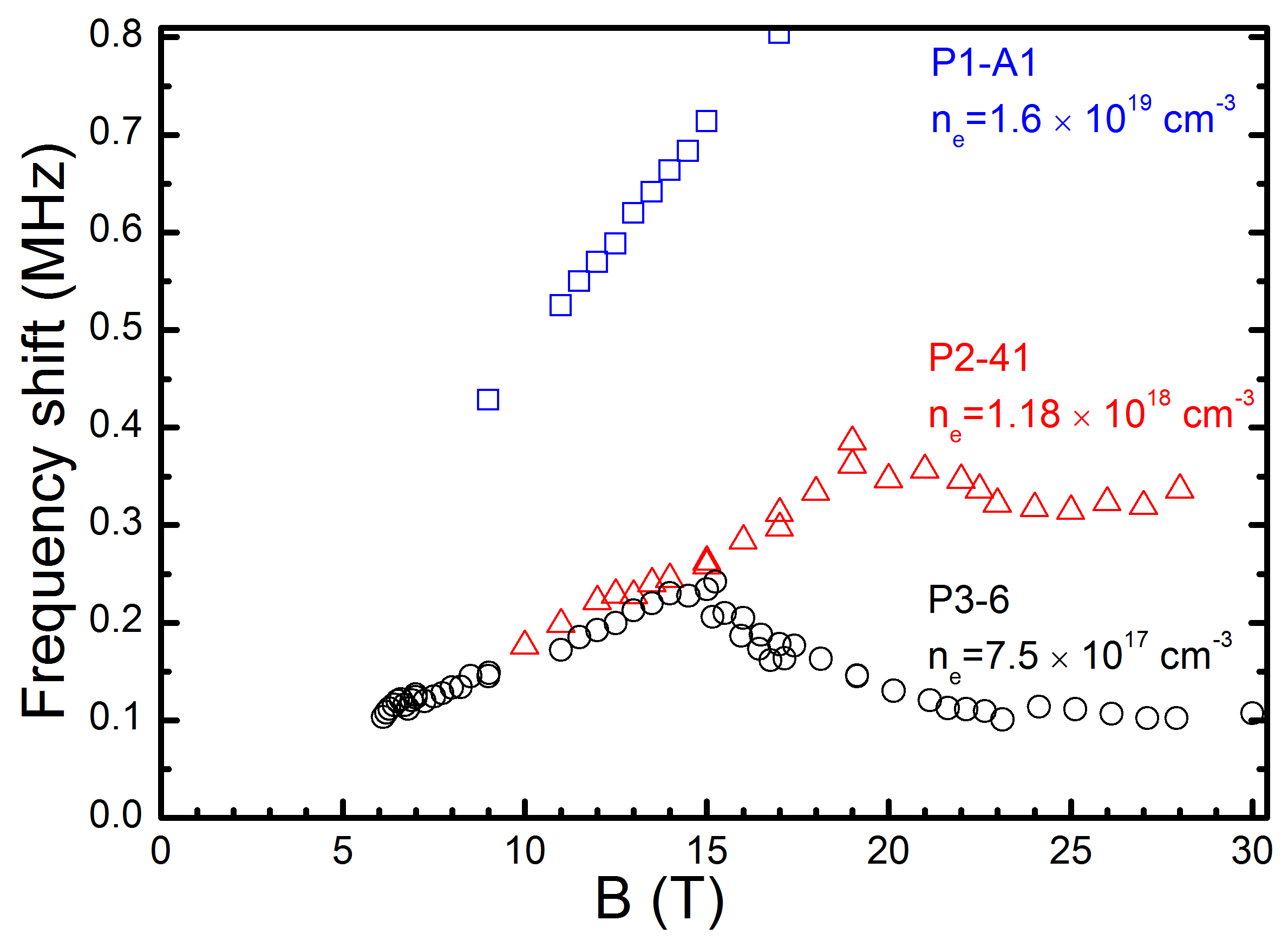}
\end{center}
\caption{(color online) NMR frequency shift ($f-f_{0}$) of the Bi
central line as a function of the magnetic field $B$ applied along
the $c$-axis, for three samples having different carrier
concentrations. $T=5$ K.} \label{Fig2}
\end{figure}

Figure~\ref{Fig1} shows a typical quadrupolar split $^{209}$Bi NMR
spectrum at 5 K and its evolution with the magnetic field $B$
parallel to the crystal $c$-axis. The spectrum consists of a
central line and four pairs of satellites due to the quadrupolar
coupling between the nuclear spin $I=\frac{9}{2}$ charge
distribution and the electric field gradients in the crystal. The
main $^{209}$Bi NMR central line is positively shifted compared to
the expected position of the bare Bi resonance. In metals and
semiconductor, this shift includes the ``Knight shift'', due to
the interaction between the free carriers and the nuclei, which is
our main focus of interest in this paper. As can be seen in
figure~\ref{Fig1}, both the central line and its satellites shift
to higher frequencies as the magnetic field increases up to $\sim$
15T, where they show a sudden drop in frequency. The field
dependence of the frequency shift ($f-f_{0}$) of the $^{209}$Bi
NMR central line observed in figure~\ref{Fig1} is plotted in
figure~\ref{Fig2} for three different samples.

At low magnetic field, the shift increases (approximatively)
linearly for all samples, with an average slope of about 0.0165
MHz/T ($\sim$ 2410 ppm) for sample P3-6. At a fixed magnetic
field, it increases with the electron density \cite{SM.II.D.}. For
samples P3-6 and P2-41 a pronounced kink appears in the NMR shift
for magnetic fields of about 15 T and 21 T, respectively. For the
high density sample (P1-A1), no saturation is observed in the
studied field range. These simple observations of density
dependent features in the shift points towards the influence of
the conduction electrons. In the simplest approach, the hyperfine
contact term gives rise to an effective magnetic field
proportional to the total number of free carriers $n_{e}$ and
their spin polarization $P=(n_{\uparrow}-n_{\downarrow})/n_{e}$,
where $n_{\uparrow}$ and $n_{\downarrow}$ are the number of spin
up and down in the system, respectively. This leads to an NMR
frequency shift which is directly connected to the spin
polarization of the system. In a bulk semiconductor the spin
polarization is primarily determined by the ratio between the
Zeeman and Fermi energies and increases with the magnetic field.
In sufficiently high magnetic fields (or low Fermi energies), a
complete spin polarization of the system can in principle be
achieved, leading to a saturation of the internal effective
magnetic field seen by the nucleus, and thus a saturation of the
NMR frequency shift. This joint condition of a low Fermi energy
and a high Zeeman energy can however only be experimentally
realized in dilute 3D electronic systems under relatively high
magnetic fields. To our knowledge, a saturation of the NMR Knight
shift has only been observed in InSb so far
~\cite{Miranda1974,Willig1972,Tunstall1988}, where the effective
electron $g$-factor is about -51.

In Bi$_2$Se$_3$, the bulk magneto-transport properties have been
successfully described in the pioneering work of K\"{o}hler
\cite{Kolherg1975}, using a model of a 3D electron gas with an
energy-dependent anisotropic Fermi surface and a spin splitting
described by an effective g-factor $g^{*}_{eff}$ = 32$ \pm3$. This
phenomenological approach was confirmed by our magneto-transport
measurements \cite{SM.I.}. A saturation of the Bi$_2$Se$_3$ Knight
shift is thus expected for a magnetic field $B_{s}$, defined in
the phenomenological approach by
$g^{*}_{eff}\mu_{B}B_{s}={2}^{\frac{2}{3}} E_{F}$, where $E_{F}$
is the zero-field Fermi energy \cite{noteSM.III.A.B.}. For samples
P3-6 and P2-41, our transport experiments yield $E_{F}$= 19.7 and
26.4 meV respectively, which with $g^{*}_{eff}$=32.5 gives
corresponding values of $B_{s}$ = 16.6 and 22.2 T, close to the
observed kinks in figure \ref{Fig2}. The absence of any deviation
of the shift in the higher carrier concentration sample is also
expected since the much higher Fermi energy pushes $B_{s}$ far
outside of our observable range. These very simple arguments
points toward a direct connection between the electronic spin
polarization and the NMR Knight shift in Bi$_{2}$Se$_{3}$.

\begin{figure}[!h]
\begin{center}
\includegraphics[width= 8.5cm]{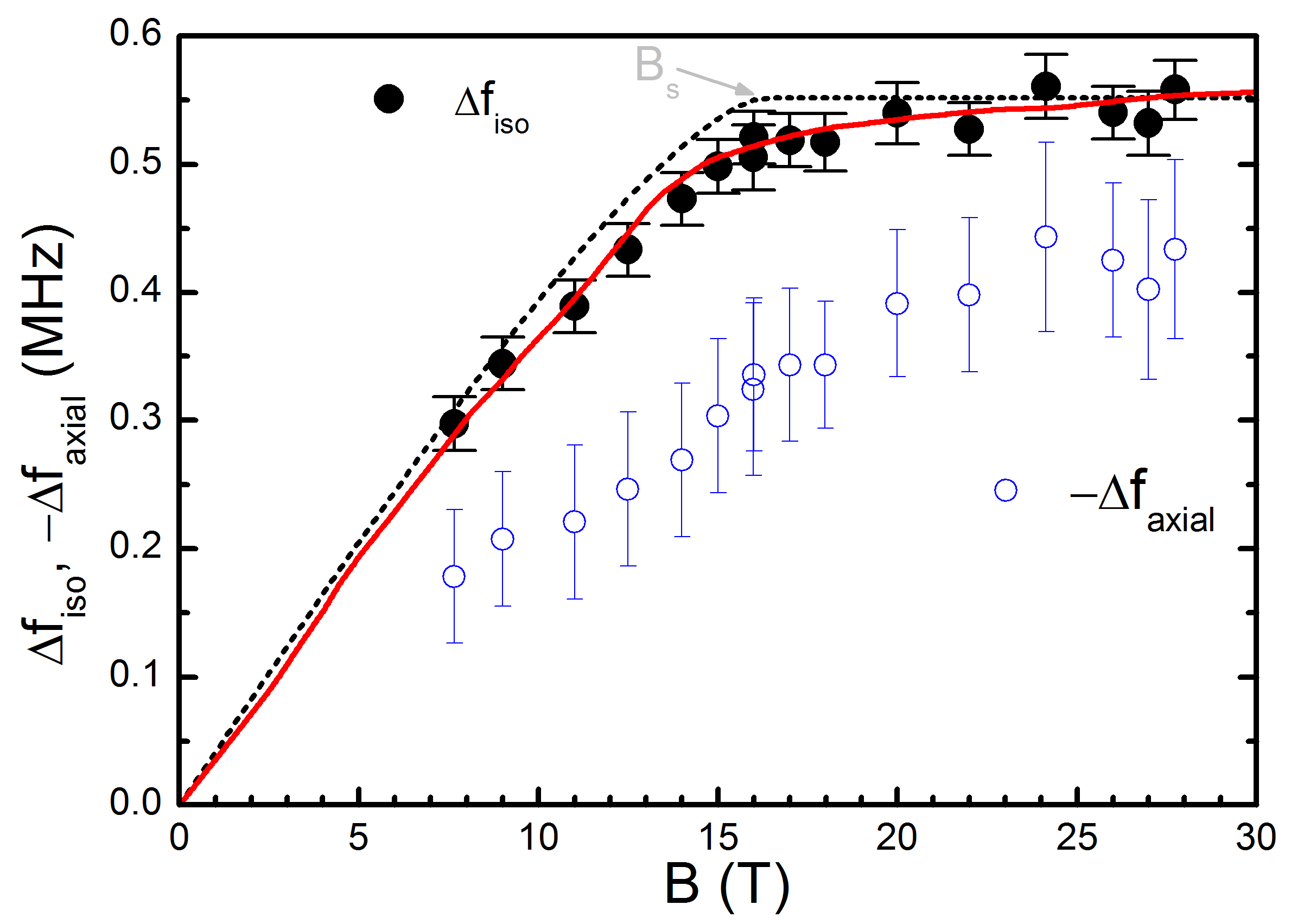}
\end{center}
\caption{(color online) Decomposition of the total frequency shift
into isotropic ($\Delta f_{iso}$) and axial parts ($\Delta
f_{axial}$) (see text) for sample P3-6. To compare the two
components on the same graph, $-\Delta f_{axial}$ is plotted. The
black dotted line is a simulations in a 3D electron gas model
($B_{s}$ indicates the field corresponding to full spin
polarization). The red solid line further includes effects of the
Landau quantization and of disorder.} \label{Fig3}
\end{figure}

We note, however, that for sample P3-6 a clear saturation of the
Knight shift to a maximum value is not observed: the shift rather
slowly drops down and appears to saturate to a lower value above
22 T. This behavior has been confirmed for different samples of
similar density and is also to some extent present for sample
P2-41. For a deeper understanding of our data, one should consider
the general expression of the hyperfine coupling Hamiltonian in
metals or doped-semiconductors ~\cite{Abragam1985}:

\begin{equation}\label{eq1}
\mathcal{H}_{HF}
=\frac{\mu_{0}}{4\pi}g_{0}\mu_{B}\gamma\hbar\textbf{I}\cdot
\left[\frac{8\pi}{3}\textbf{S}\delta (\textbf{r})
-(\frac{\textbf{S}} {r^3} - \frac{3(\textbf{S}\cdot
\textbf{r})\textbf{r}}{r^5})+ \frac{\textbf{l}}{r^3} \right],
\end{equation}
where $g_{0}$ and $\mu_{B}$ are the electronic bare $g$-factor and
Bohr magneton, respectively, $\gamma$ is the nuclear gyromagnetic
ratio, \textbf{I} and \textbf{S} are the nuclear and electronic
spin respectively, \textbf{r} is the electron-nuclei position
vector and \textbf{l} the orbital momentum of the conduction
electron. The first term in the bracket is the ``contact''
hyperfine interaction, which originates from the density
probability of the electron exactly at the nuclear site ($r=0$).
This coupling is isotropic and dominates the hyperfine interaction
in the case of an \textit{s}-like electronic wave function. This
yields a direct access to the electronic spin polarization (see
e.g. Ref.~\onlinecite{Tiemann2012,Piot2012}). The second term is
anisotropic and results from the dipolar coupling between nuclear
and electronic moment away from the $\textbf{r}=0$ singularity,
which dominates in the case of \textit{p}-like electronic wave
function (see e.g. Ref.~\onlinecite{Fischer2008,Chekhovich2013}).
The last term is related to the orbital momentum of the conduction
electrons. In solids with non-degenerate orbital levels (which is
the case for Bi$_2$Se$_3$ \cite{Zhangtheo2009,Cao2013}), the
average orbital momentum $<\textbf{l}>=0$, and this term usually
only gives rise to a second order contribution \cite{Winter1971}.
However, in the presence of a strong spin-orbit coupling, a direct
orbital contribution to the Knight shift may exist and be
non-negligible: an example is the case of \textit{p}-like
electronic wave functions in $n$-type PbTe
\cite{Leloup1973,Hota1991}. As for all the terms in Eq.\ref{eq1},
the relevance of this contribution is strongly dependent on the
nature of the wave function, and no theoretical estimations of its
strength in Bi$_2$Se$_3$ have so far been proposed
\cite{orbitalshiftnote}. Very recently, the presence of both
anisotropic and isotropic contributions in the total NMR shift in
Bi$_2$Se$_3$ has been reported at low carrier density
\cite{Young2012}. To further investigate the origin of the
observed downturn in high magnetic fields (Fig.\ref{Fig2}), we
have performed angular dependent measurements of the NMR shift.
For the axial symmetry, present in Bi$_2$Se$_3$, one can decompose
the total frequency shift $\Delta f$ into two contributions, an
isotropic contribution $\Delta f_{iso}$ and an anisotropic one
$\Delta f_{axial}$, by writing $\Delta f=\Delta f_{iso}+\Delta
f_{axial}$. For a given angle $\theta$ between the magnetic field
$B$ and the $c$-axis, the total shift can then be written with an
explicit angular dependence as $\Delta
f=[\mathcal{A}_{iso}+(3\cos^2\theta-1)\mathcal{A}_{axial}] n_{e}
\times P(\theta)$. $\mathcal{A}_{iso}$ and $\mathcal{A}_{axial}$
are the isotropic and anisotropic hyperfine coupling constants,
respectively, and the spin polarization $P$ may depend on $\theta$
in the case of anisotropic conduction band parameters. The
isotropic and the axial part of the shift can be extracted from
the rotational pattern of the central line \cite{SM.II.B.2}.
Figure~\ref{Fig3} shows the isotropic and axial components of the
frequency shift for sample P3-6.

A clear saturation of the isotropic component of the Knight shift
is observed above 16 T, and can be attributed, as discussed above,
to the full spin polarization of the electronic system manifesting
itself directly through the ``contact'' term of the hyperfine
coupling. The frequency shift can be calculated by writing $\Delta
f_{iso}=\mathcal{A}_{iso}\times n_{e} \times P$, where the spin
polarization $P$ is calculated in the simple 3D electron gas model
introduced above. The two free parameters of this model,
$g^{*}_{eff}$ and $\mathcal{A}_{iso}$, can be determined
independently. As the Fermi energy is precisely estimated from
transport measurements, $g^{*}_{eff}$ is only determined by the
position of the saturating field $B_{s}$. $\mathcal{A}_{iso}$ is
then adjusted to reproduce the frequency shift value, from the low
field region to the saturation. The result of this simulation is
plotted as a dotted line in figure \ref{Fig3} and catches the
observed saturation very well \cite{SM.III.A.B.}. An even better
agreement with the experimental data can further be obtained by
incorporating additional elements to our model, such as the Landau
quantization of the electronic density of states (fully
characterized from transport experiments \cite{SM.I.}) and the
effect of disorder which reduces the spin polarization through a
level broadening. The result is plotted as solid line in figure
\ref{Fig3} and gives a perfect quantitative description of our
experimental data. Small modulations of the Knight shift can even
be observed and attributed to the effect of the Landau
quantization on the spin polarization, see e.g. the shallow
downturn around 10.8 T. In the case of Bi$_2$Se$_3$ the
characteristics of these oscillations reflect, in particular, the
peculiar commensurability of the spin and cyclotron gap
\cite{SM.I.}, expected within the generalized 3D Dirac Hamiltonian
for massive fermions \cite{Orlita}. The obtained effective
electron $g$-factor is $g^{*}_{eff} = + 32.8 \pm 1$, in a very
good agreement with transport experiments performed on the same
samples \cite{SM.I.}, and the previously reported value of 32$
\pm3$ \cite{Kolherg1975}. We note that, in contrast to transport
experiments, NMR experiments unambiguously determines the sign of
the $g$-factor which directly impacts the sign of the Knight
shift. The extracted value of the isotropic hyperfine coupling
constant $\mathcal{A}_{iso}=6.94 \pm 0.16 \times 10^{-13}
cm^{3}/s$ enables us to estimate the density of the electronic
wave function at the nuclear site
$d_{\textrm{Bi}}=8.72\times10^{25} cm^{-3}$ \cite{SM.II.B.3.}.

The obtained values of $\mathcal{A}_{iso}$ and $d_{\textrm{Bi}}$
are comparable to or even higher than values determined for GaAs
electron systems exhibiting pure \textit{s}-like wave function
(e.g. $\mathcal{A}_{iso}=4.5 \pm 0.2 \times 10^{-13} cm^{3}/s$
\cite{Kuzma1998} and $d_{\textrm{Ga}} = 5.8 \times 10^{25} {\rm
cm}^{-3}$ \cite{Paget1977} for $^{71}$Ga in GaAs). This is at
first sight surprising, since from a low energy expansion around
the Dirac point \cite{Zhangtheo2009}, it is expected that the wave
function describing the Dirac surface states, as well as the one
describing the bulk conduction band minimum (CBM) and bulk valence
band maximum (VBM) states, is of \textit{p}-wave nature. In this
case, the NMR shift produced by conduction electrons should be, at
least at low energy, purely anisotropic. Our results, showing a
very large isotropic component in the Knight shift for Fermi
energies as low as 18 meV, therefore cannot be understood within
this approach. They suggest that a non-negligible portions of the
electronic wavefunction is of \textit{s}-wave nature. In fact, our
results could be consistent with very recent theoretical results
\cite{Pertsova2013} showing that about 10 \% of the electronic
wave function at the Bi sites are of s-wave nature, emerging from
deeper Bi ``\textit{6s}'' orbitals taking part in shaping the wave
function both in the VBM and CBM. Another interesting result of
these calculations is the apparent absence of an \textit{s}-like
wave function at the Se site: this is consistent with the smaller
shift observed at the same magnetic field for the $^{77}$Se NMR
\cite{SM.II.C.} and the extremely long relaxation time \cite{SM.II.C.,Taylor2012}, showing a weaker coupling of the Se nuclei to the conduction electrons.
 The remaining dominant \textit{p}-wave nature of these wave
functions leads to the anisotropic dipolar term in the $^{209}$Bi
NMR shift \cite{SM.II.B.4.}. Its magnetic field behavior turns out
to be more complex than the one of the isotropic component, as can
be seen in figure \ref{Fig3}. This illustrates the importance of
having a sizeable isotropic component in the Knight shift as a
reliable and direct access to the electronic spin polarization. We
finally note that our temperature-dependent studies of the Knight
shift \cite{SM.III.E.}, constitute an additional manifestation of
the direct connection between the NMR shift and the electronic
spin polarization and confirms our general approach.

In conclusion, combined high magnetic fields NMR and transport
measurements have enabled us to evidence a direct link between the
electron spin degree of freedom and the nuclear system, for the
first time in a material of the topological insulator class. We
precisely estimate the ``contact term'' of the hyperfine coupling
and determine the amplitude and sign of an effective electronic
$g$-factor phenomenologically describing the spin splitting in
Bi$_{2}$Se$_{3}$. Our observation of a large isotropic NMR shift
reveals a non-negligible proportion of \textit{s}-like electronic
states near the bulk band gap of this topological insulator.

This work was supported by the European Community under contract
EC-EuroMagNetII-228043. M.O, M.P. and B.A.P acknowledge support
from the ERC-2012-AdG-320590 MOMB grant. Work in Poland has been
supported by the research project: NCN UMO - 2011/03/B/ST3/03362
Polska.


\begin{thebibliography}{41}
\bibitem[{orb()}]{orbitalshiftnote}
\bibinfo{note}{We note that our study of the temperature dependence of the
  shift \cite{SM.III.E.} suggests that, in our system, a spin-independent
  orbital contribution can be neglected in front of the spin-dependent ones.}
\end{thebibliography}
\end{document}


\title{Hyperfine coupling and spin polarization in the bulk of topological insulator Bi$_{2}$Se$_{3}$: SUPPLEMENTARY INFORMATION}

\maketitle

This supplemental material is organized in three main sections.
Section I discusses the transport properties and conduction band
parameters in Bi$_{2}$Se$_{3}$, section II formally presents the
model used for our spin polarization and Knight shift simulations,
and section III presents further details on our NMR experiments.

\section{Transport properties and conduction band parameters}

\subsection{Sample characteristics}

Magneto transport experiments were conducted on Bi$_{2}$Se$_{3}$
sample slices on which silver paste contacts were deposited in a
``Hall bar-like'' or Van der Pauw configuration. The measurements
were performed using a standard low frequency lock-in technique in
a variable temperature insert for temperature between 1.2 and 40
K, and in magnetic fields up to 30 T produced by a 20 MW resistive
magnet. The characteristics of the studied sample are summarized
in table \ref{tab1}, together with parameters extracted from the
transport measurement analysis.

\begin{table}[h]
\begin{center}
\begin{tabular}{ccccccccc}
\hline\hline &$\varepsilon$ ($\mu$m) & T$_{D}$(K)& $F$(T) [$F_{\|}$(T)] &$m^{*}( m_{e}$) [m$^{*}_{\|}( m_{e})$] & $E_{F}$ (meV) & $n_{e}$ ($cm^{-3}$)& \\
\hline

P2-6A & 60 & 11.2 \footnote{for comparison, the corresponding 4K mobility is $\mu=0.665$ $m^{2}/Vs$} & 19.6 [28.8] & 0.123$\pm$0.005 [0.182$\pm$0.003] & 18.5& 6.76$\times 10^{17}$\\
\textbf{P2-6B} & 60  & 9.6 & 21.2 & - & 20.0 & 7.65$\times 10^{17}$ \\
\textbf{P3-6} & 120  &9.6 & 20.9 & - & 19.7& 7.47$\times 10^{17}$ \\
\textbf{P2-41} & 10  &  15 & 28.0 [42.6] & 0.123$\pm$0.007 [0.206$\pm$0.002] & 26.4 & 1.17$\times 10^{18}$\\
P1-3 & 10  & 17.4  &  154.8 & 0.1307$\pm$0.002 & 137& 1.80$\times 10^{19}$\\
\textbf{P1-A1} & 160  & -  & 155.5 & - & 137.9 & 1.81$\times 10^{19}$ \\

\end{tabular}
\end{center}
\caption{Characteristics and physical parameters of the
Bi$_{2}$Se$_{3}$ samples: thickness $\varepsilon$, Dingle
temperature T$_{D}$(K), main frequency of the quantum oscillations
$F$(T) (frequency $F_{\|}$ for $B \perp$ \textit{c}-axis indicated
in brackets when available), effective mass $m^{*}$ in units of
the rest electron mass $m_{e}$ (m$^{*}_{\|}$ for $B \perp$
\textit{c}-axis indicated in brackets when available), Fermi
energy $E_{F}$, and electron density $n_{e}$. The samples used for
both transport and NMR measurements are highlighted in bold
case.}\label{tab1}
\end{table}

\begin{figure}[h]
\begin{center}
\includegraphics[width=15cm]{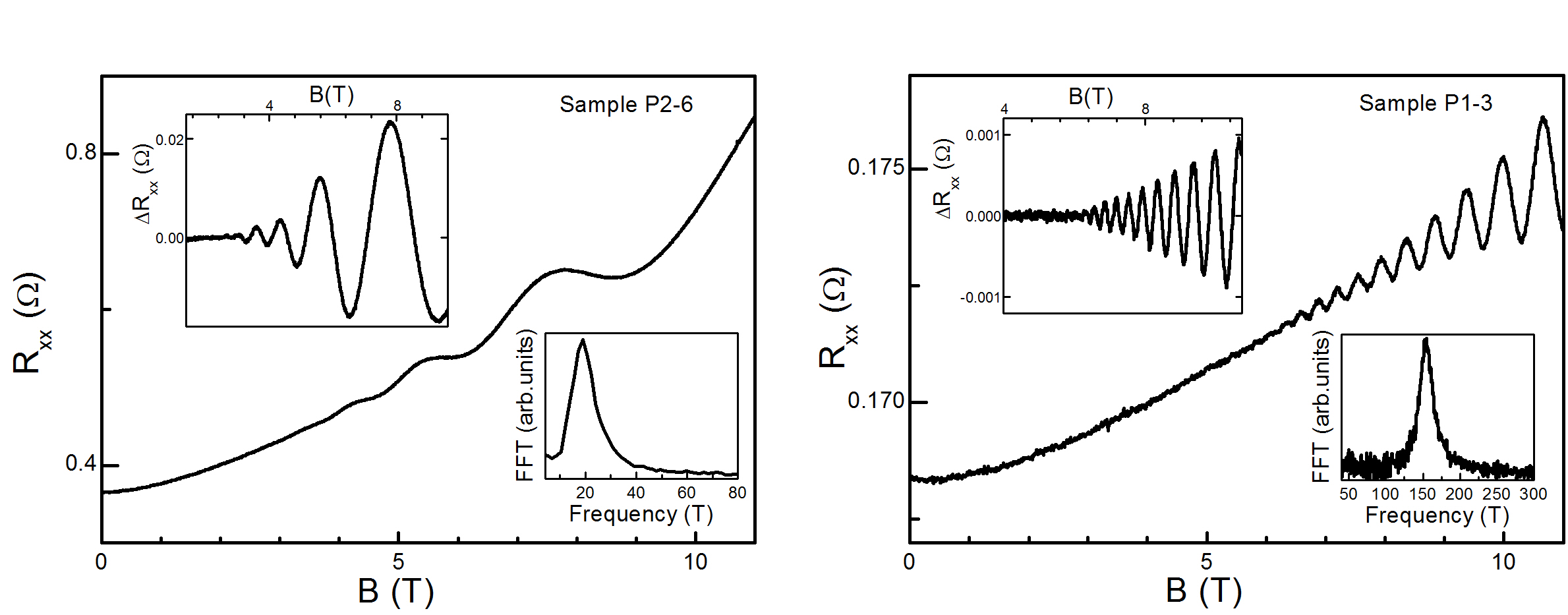}
\end{center}
\caption{Longitudinal resistance $R_{xx}$ versus total magnetic
field $B \parallel$ \textit{c}-axis for sample P2-6 (left panel)
and P1-3 (right panel) at T=1.4 K. Upper left insets: oscillatory
resistance $\Delta R_{xx}$. Lower right insets: Fourier transform
of $\Delta R_{xx}(1/B)$.}\label{Fig1SI}
\end{figure}

The Shubnikov-De Haas oscillations (SdH) were observed in the low
temperature magneto-transport and show, as usually observed, a
single frequency connected to the single extremal cross section
between the magnetic orbits and the Fermi surface. Representative
results are shown for extremal values of our carrier density range
in figure S\ref{Fig1SI}. The longitudinal resistance $R_{xx}$,
symmetrized for positive and negative magnetic fields, shows
quantum oscillations superimposed on a monotonously increasing
background. The background can be removed either by subtracting a
polynomial function to the data or by ``adjacent averaging''
method, and the corresponding oscillatory resistance is plotted in
the upper left insets.

The sample quality was characterized by the determination of the
Dingle temperature extracted from the field damping of the SdH
oscillations. This was done using the standard Lifshitz-Kosevich
formalism \cite{LK56} at the lowest temperatures. The resulting
Dingle temperatures, reported in table \ref{tab1}, reflect the
typical trend of a disorder increasing with the carrier
concentration, related to a higher concentration of Se vacancies
in highly doped samples. The Hall voltage was also recorded to
estimate the carrier density and, consistently with previous
reports, \cite{kohlerCB1973} an offset between the ``Hall
density'' and the ``SdH density'' was observed with increasing
carrier density. Because of geometrical errors possibly occurring
in the determination of the electron density via the Hall voltage
(particularly in anisotropic materials), quantum oscillations were
considered to be a more reliable way of determining the carrier
density (see also I.B.3. and III.D).

\subsection{Fermi surface anisotropy}

\subsubsection{Angular dependence of the quantum oscillations}

\begin{figure}[!h]
\begin{center}
\includegraphics[width=15cm]{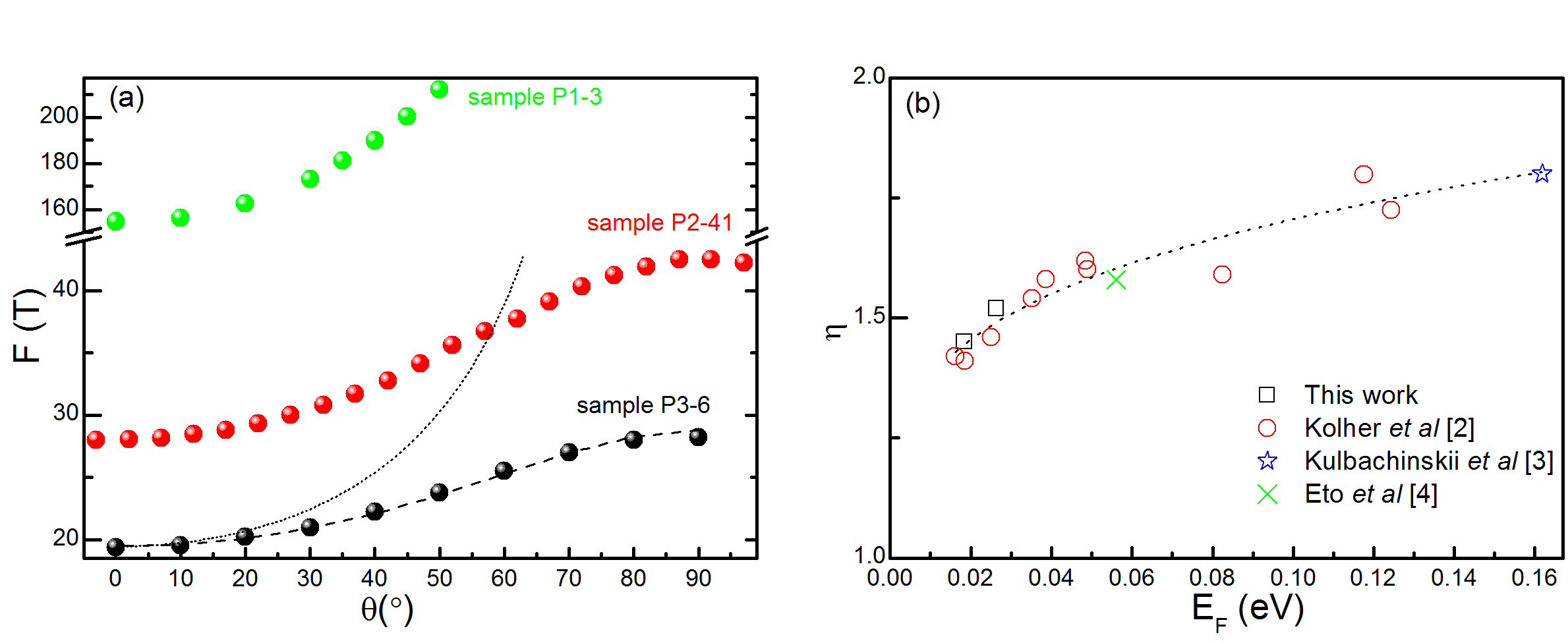}
\end{center}
\caption{(color online)(a) SdH oscillations frequency F as a
function of the tilt angle $\theta$ for three samples with
different carrier concentration (increasing from P3-6 to P2-41 to
P1-3, respectively, from bottom to top). Calculated angular
dependence of F for a 3D ellipsoidal Fermi surface (dashed-line),
and for a 2D system (dotted-line). (b) Anisotropy $\eta=F_{\|}/F$
as a function of the Fermi energy in the conduction band,
including previously published results. The dotted-line is a
power-law fit of the data.}\label{Fig2SI}
\end{figure}

The angular dependence of the quantum oscillations was studied to
map out the bulk conduction band Fermi surface for different
values of the Fermi energy. SdH measurements were performed with
an \textit{in-situ} rotation stage, enabling us to tune the angle
$\theta$ between the magnetic field and the crystal
\textit{c}-axis continuously from 0 to 90 degrees.
The $\theta=0$ data were initially symmetrized by changing the
polarity of the magnetic field to check that contact misalignment
had a negligible impact on the Fourier transform analysis. As a
double check, for a few samples the data were symmetrized for
\textit{each} angle and the Fourier transform analysis performed
on the fully symmetrized data set, showing no noticeable
difference compared to the analysis performed with non-symmetrized
data. The variation of the oscillation frequency as a function of
the tilt angle $\theta$ is reported in figure S\ref{Fig2SI}.a for
three samples with different carrier concentration (increasing
from P3-6 to P2-41 to P1-3, respectively).

Since the oscillation frequency is directly proportional to the
extremal cross section of the Fermi surface in the momentum space,
the Fermi surface can be mapped out from this angular dependence.
As can be seen in figure S\ref{Fig2SI}.a, the frequency $F$
increases as a function of $\theta$ in a way characteristic for an
ellipsoidal Fermi surface (dashed-line), consistent with
pioneering studies of the bulk conduction band at low
energy.\cite{kohlerCB1973} In this early work, the Fermi surface
was described within a simple model of massive carriers with a
parabolic (non-parabolic) dispersion in the k$_{\bot}$ (k$_{\|}$)
direction, where k$_{\bot}$ (k$_{\|}$) is the momentum in the
direction perpendicular (parallel) to the \textit{c}-axis of the
crystal. This is associated with an \textit{increasing anisotropy
of the Fermi surface observed as the Fermi level increases} in the
conduction band. Accordingly, in figure S\ref{Fig2SI}.a, the
frequency increase with $\theta$ gets steeper as the carrier
density increases (sample P3-6 to P2-41 to P1-3), confirming the
higher anisotropy of the Fermi surface at higher Fermi energies.
Another interesting aspect revealed by this data is that the cross
section can not be perfectly reproduced by assuming a purely
ellipsoidal Fermi surface above a certain Fermi energy. For
example, fitting the low $\theta$ trend observed for sample P1-3
with an ellipsoidal model for the Fermi surface would give a much
higher anisotropy than the one measured in such samples
\cite{kohlerCB1973}, suggesting that the Fermi surface is
``squashed'' along the \textit{c}-axis. The ratio $F_{\|}/F$ of
the SdH frequencies at $\theta=0^{\circ}$ and $\theta=90^{\circ}$,
which reflect the anisotropy $\eta$ of the Fermi surface
(independently of its shape), is plotted in figure S\ref{Fig2SI}.b
together with results previously reported in the literature for
samples with different carrier concentration.
\cite{kohlerCB1973,Kulba1999,Eto2010} The determination of the
Fermi energy is detailed below in section I.B.3. The dotted-line
is a power-law fit of all the combined data sets used in our work
to characterize the energy dependent anisotropy of the Fermi
surface in Bi$_{2}$Se$_{3}$.

\subsubsection{Effective mass}

\begin{figure}[!h]
\begin{center}
\includegraphics[width=9cm]{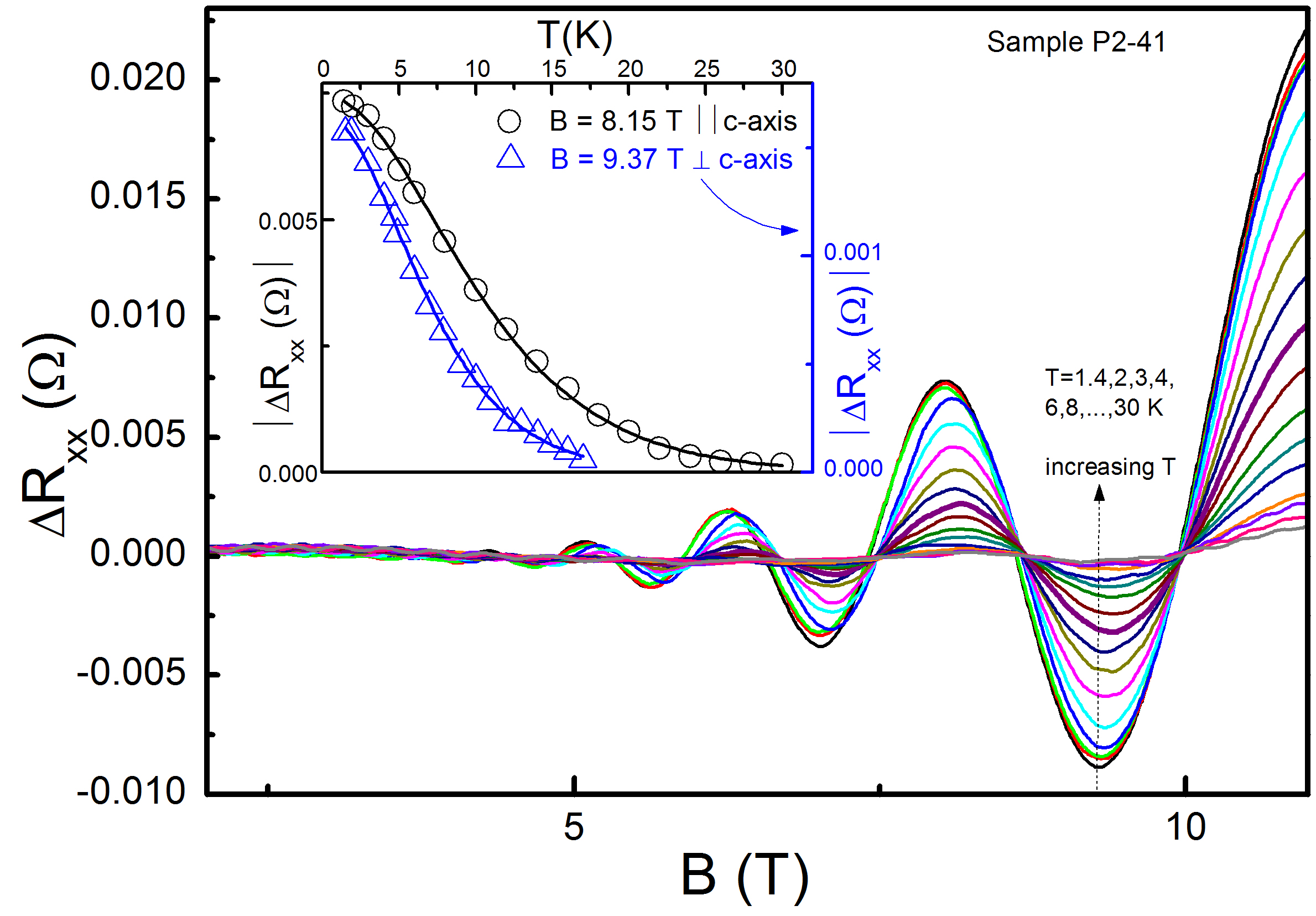}
\end{center}
\caption{(color online) Oscillatory magnetoresistance $\Delta
R_{xx}$ versus total magnetic field for different temperatures for
sample P2-41. Inset: temperature dependence of $\Delta R_{xx}$ at
the resistance extremum at B=8.15 T for $B \parallel$
\textit{c}-axis (left-axis) and at B=9.37 T for $B \perp$
\textit{c}-axis (right-axis) (magneto resistance curve not
shown).}\label{Fig3SI}
\end{figure}

The temperature dependence of the SdH oscillations was measured
for samples P2-6A, P2-41, P1-3 between 1 and 40 K for magnetic
fields up to 11 T. In figure S\ref{Fig3SI}, we report typical
results obtained on sample P2-41, for temperatures between 1.2 and
30 K.

In the  inset of figure S\ref{Fig3SI}, we plot the temperature
dependence of the oscillations amplitude at a fixed magnetic
fields for different configurations: the standard $B
\parallel$ \textit{c}-axis configuration (open circles), and the
$B \perp$ \textit{c}-axis configuration (open triangles) where the
magnetic field is in the \textit{(a,b)} plane of the crystal. The
temperature damping can be very well-described by the
Lifshitz-Kosevich formalism \cite{LK56}, operational in this case
of quantum oscillations of moderate amplitude. From this formalism
one can extract an energy gap $\Delta$ separating the quantized
energy levels, from which we can define an ``apparent'' effective
mass for massive carriers, $m^{*}=(\hbar eB)/ \Delta$. The values
reported in table \ref{tab1} are the one measured at the lowest
possible magnetic fields, and were found to be almost constant
over the field range studied. The effective mass in the $B \perp$
\textit{c}-axis configuration, $m^{*}_{\parallel}$, is also
reported in table \ref{tab1}. As can be seen, $m^{*}_{\parallel}$
is larger than $m^{*}$ and increases with the Fermi energy, while
$m^{*}$ is approximatively energy independent in the studied
range. This behavior is related to the energy-dependent anisotropy
of the Fermi surface discussed in the previous section. In the
case of a perfect ellipsoid, the ratio $m^{*}_{\parallel}/m^{*}$
is equal to the cross section ratio $F_{\|}/F$. The small
differences between these two ratios for sample P2-41 shows the
ellipsoidal model starts to breakdown already for Fermi energies
above 25 meV.

\subsubsection{Fermi energy and carrier concentration}

From the frequency $F$ of the SdH oscillations, one can deduce the
extremal Fermi surface cross section in the momentum space,
$CS=\pi k_{F}^{2}$ (where $k_{F}$ is the Fermi wave vector), given
by $CS=(\hbar F)/(2\pi e)$. As argued above, an energy-independent
effective mass is observed in the $B \parallel c$-axis
configuration. One can therefore connect the Fermi energy $E_{F}$
to the extremal Fermi surface cross section $CS$ through the
measured effective mass $m^{*}$, by writing
$E_{F}=(\hbar^{2}CS)/(2 \pi m^{*})$. This enables us to determine
the Fermi energies reported in table \ref{tab1}, which define the
energy range probed in the conduction band in our experiment. The
carrier density can then be calculated from the Fermi energy and
the density of states (DOS) of the system. In such non-parabolic
systems, the DOS has to be derived numerically from the actual
dispersion relation. To simplify the problem, we approximate the
DOS by introducing an energy-dependent effective mass along the
\textit{c}-axis, which aims at reproducing phenomenologically the
energy-dependent anisotropy of the Fermi surface $\eta(E)$. The
corresponding mass $m^{*}_{z}$ is given in an ellipsoidal Fermi
surface approximation by $m^{*}_{z}=m^{*}(\eta(E))^{2}$, where
$\eta(E)=1.32E^{0.272}+1$ is determined from the fit of the data
in figure S\ref{Fig2SI} (dotted-line).

The spin degenerate density of states $D_{3\textrm{D}}(E)$ of the
3D Fermi sea is then approximated by:

\begin{equation}\label{DOS}
D_{3\textrm{D}}(E)=\frac{1}{2\pi^{2}}\sqrt{\frac{2m^{*}_{z}}{\hbar^{2}}}\frac{2m^{*}}{\hbar^{2}}\sqrt{E}
\end{equation}

The carrier density $n_{e}$ is then:

\begin{equation}\label{dens}
n_{e} = \int_{-\infty}^{E_{F}} D_{3\textrm{D}}(E) dE,
\end{equation}

The electron densities obtained are reported in table \ref{tab1}.
For the samples in which the effective mass was not directly
measured, we have taken the effective mass value obtained for
samples exhibiting a very similar oscillation frequency.

\subsection{High magnetic field transport and Zeeman energy}

The high magnetic field magneto transport was measured up to 30 T
for samples P2-6B and P3-6, and is very similar to the data
previously published in Ref.~\onlinecite{Kolherg1975}. In
particular, the observed minima and maxima in the quantum
oscillations above $B=15$ T demonstrate that two Landau levels are
below the Fermi energy at $B=15$ T. This, together with the
absence of spin splitting in the quantum oscillation up to 30T, is
fully consistent with the commensurability observed between the
Zeeman and the cyclotron energies in \textit{n}-type
Bi$_{2}$Se$_{3}$. \cite{Kolherg1975,Kulba1999}

By simulating the high field quantum oscillations with the same
3DEG model as the one used in Ref.~\onlinecite{Kolherg1975}, with
the parameters $m^{*}$ and $T_{D}$ experimentally determined for
our samples, one can draw the conclusion that the ratio $M$
between the Zeeman and cyclotron energies is:

\begin{equation}\label{eqcoincidence}
M = \frac{g^{*}_{eff}\mu_{B}B}{\hbar \frac{e B}{m^{*}}} = 2.0 \pm
0.2
\end{equation}

This gives an estimation of the electronic $g$-factor
$g^{*}_{eff}=32.5 \pm 3$ for sample P2-6B and P3-6, in agreement
with Ref.~\onlinecite{Kolherg1975}. We note that from the study of
the phase of the quantum oscillations at high magnetic fields, the
condition $M=1$ can be ruled out. The magnetic field dependence of
the Fermi energy in the presence of a high Zeeman energy is also
evidenced by a significant deviation of the phase of the quantum
oscillations, as theoretically expected (see
Ref.~\onlinecite{Shoenberg1984}). This should be taken into
account with great care when analyzing the phase of quantum
oscillations \textit{to search for 2D surface carriers in
Bi$_{2}$Se$_{3}$ systems}. Finally, the SdH phase analysis
performed in the $B \perp $ \textit{c}-axis configuration up to 11
T enables us to evidence an anisotropy of the $g$-factor. For
example, for sample P2-41 the $g$-factor in the crystal
\textit{(a,b)} plane, $g^{*}_{eff\perp}$, is estimated to be
$g^{*}_{eff}\perp = 19 \pm 2$, whereas $g^{*}_{eff}=31 \pm 2$
along the \textit{c}-axis. A similar anisotropy of the g-factor
was also observed in recent spin resonance experiments at low
magnetic fields \cite{Wolos2013}. This g-factor anisotropy,
together with the previously discussed effective mass anisotropy,
leads to the fulfillment of the coincidence condition $M \sim 2$
independently of the magnetic field orientation, in agreement with
Ref.~\onlinecite{Kolherg1975}. We finally note that the success of
this phenomenological approach, in which the spin gap
$g^{*}_{eff}\mu_{B}B$ turns out to equal \textit{twice} the value
of the cyclotron gap associated with a parabolic in-plane
dispersion, is not random but rooted to the 3D Dirac Hamiltonian
for massive fermions applied to topological insulators with low
electron-hole asymmetry and roughly parabolic bands. \cite{Orlita}

\clearpage

\section{Knight shift and spin polarisation simulations}

\subsection{Polarization of a 3D electron gas}

The simplest version of the model used to simulate the NMR Knight
shift is based on the calculation of the electronic spin
polarization of a 3D Fermi sea, characterized by an effective
$g$-factor $g^{*}_{eff}$ defining the total spin splitting
$g^{*}_{eff}\mu_{B}B$. One should note that the $g$-factor is
different from the bare Zeeman $g$-factor since it includes the
effect of the spin-orbit coupling to define an effective spin gap.
As the magnetic field is turned on, the spin degenerate density of
states of the 3D Fermi sea given by Eq. \ref{DOS} is split by the
Zeeman energy. The corresponding spin-up and spin-down populations
are given by:

\begin{equation}\label{densspins}
n_{\uparrow,\downarrow} = \frac{1}{2} \int_{\mp
(g^{*}_{eff}\mu_{B}B)/2}^{E_{F}} D_{3\textrm{D}}(E\pm
g^{*}_{eff}\mu_{B}B/2) dE,
\end{equation}

and the Fermi energy is defined from the total electronic density
$n_{e}$:

\begin{equation}\label{dens3DEG}
n_{e} = n_{\uparrow}(E_{F},B)+ n_{\downarrow}(E_{F},B)
\end{equation}

This results in a field-dependant Fermi energy, which slowly
decreases with the field until the point of full spin polarization
is reached. At this point the Fermi energy follows the faster
linear-in-$B$ decrease of the lower (fully occupied) spin branch
of the density of states. This rapid motion of the Fermi energy
with the magnetic field can be directly observed in the peak
position of the SdH oscillations at high magnetic field (see also
section I.C). The spin polarization under magnetic field is
calculated using:

\begin{equation}\label{pola3DEG}
P=(n_{\uparrow}-n_{\downarrow})/n_{e}
\end{equation}

The frequency Knight shift $\Delta f_{iso}$ is then calculated as
described in the main text by using $\Delta
f_{iso}=\mathcal{A}_{iso}\times n_{e} \times P$, where
$\mathcal{A}_{iso}$ is the (adjustable) hyperfine coupling
constant.

\subsection{Landau quantization and disorder}

In bulk semiconductors or metal subjected to an intense magnetic
field, the electron kinetic energy is modified by the Lorentz
force,  which leads to the formation of Landau bands originating
from the intersection of the electron cyclotron orbits with the 3D
Fermi surface. Assuming Lorentzian-shaped Landau levels (LL), the
resulting ``spin up'' and ``spin down'' density of states can be
written:

\begin{equation}\label{DOSLLL}
D_{\textrm{LL} \uparrow,\downarrow}= \frac{1}{2}
\frac{\sqrt{m^{*}_{z}}}{2\pi^{2}\hbar^{2}}eB\sum_{N_{\uparrow,\downarrow}=0}^{N_{max}}
\left[\frac{(E-E_{\textrm{N}
\uparrow,\downarrow})+\sqrt{(E-E_{\textrm{N}
\uparrow,\downarrow})^{2}+\Gamma^{2}}}{(E-E_{\textrm{N}
\uparrow,\downarrow})^{2}+\Gamma^{2}}\right]^{0.5} ,
\end{equation}

where $E_{\textrm{N}}$ are the LL eigen energy given by : $
E_{\textrm{N} \uparrow,\downarrow}=(N+\frac{1}{2})\hbar \frac{e
B}{m^{*}} \pm g^{*}_{eff}\mu_{B}B $ for the ``spin up'' (``spin
down'') spin subbands. $\hbar \frac{e B}{m^{*}}$ is the cyclotron
gap between Landau levels. $N$ is the Landau level index (a
positive integer), and $N_{max}$ is the Landau level index of the
highest energy occupied LL. $\Gamma$ is the Landau level
broadening related to the Dingle temperature $T_{D}$ by $\Gamma=
\pi k_{B} T_{D}$, where $k_{B}$ is the Boltzmann constant. For
each sample, $\Gamma$ is a therefore known and imposed from the
SdH oscillation analysis. $m^{*}_{z}$ is still taking into account
the energy dependant Fermi surface anisotropy as described above.

The Fermi energy is then obtained by solving the following
equation :

\begin{equation}\label{dens3DEGLL}
n_{e} = \int_{(\frac{1}{2})\hbar \frac{e B}{m^{*}} +
g^{*}_{eff}\mu_{B}B-100\Gamma}^{E_{F}}
\left(D_{\textrm{LL}\uparrow}(E)+D_{\textrm{LL}\downarrow}(E)\right)
dE,
\end{equation}

and the spin polarization is:

\begin{equation}\label{pola3DEGLL}
P= \frac{1}{n_{e}} \int_{(\frac{1}{2})\hbar \frac{e B}{m^{*}} -
g^{*}_{eff}\mu_{B}B-100\Gamma}^{E_{F}}
\left(D_{\textrm{LL}\uparrow}(E) -
D_{\textrm{LL}\downarrow}(E)\right) dE ,
\end{equation}

The presence of the Landau bands creates an energy modulation of
the 3D density of states which become important when the cyclotron
gap exceeds the LL disorder broadening $\Gamma$. In the presence
of a non-zero Zeeman energy, this can lead to an additional
variation of the spin polarization in a magnetic field due to a
different density of ``spin down'' and ``spin up'' states at the
Fermi level, which superimposes on the usual ``Zeeman-induced
polarization''. A particularity of Bi$_2$Se$_3$, which plays an
important role in the determination of the behavior of the spin
polarization, is the commensurability between spin and cyclotron
gap ($(g^{*}_{eff}\mu_{B}B)/(\hbar \frac{e B}{m^{*}}) \sim 2$, see
section I.C). This leads to a coincidence in the energy positions
of the Landau levels $N_{\uparrow}$ and $(N+2)_{\downarrow}$,
which tends to weaken the polarization oscillations because of the
similarities of the ``spin down'' and ``spin up'' density of
states at the Fermi level. However, in contrast with 2D systems
where such perfect coincidence would lead to a quenching of the
polarization oscillations in a magnetic field, in a 3D system some
oscillations are still expected because the density of states at
the Fermi level also involves the lowest Landau bands. These small
oscillations are seen both in our calculations and experimental
data in figure 3 in the main text. When the magnetic field reaches
the value corresponding to the full spin polarization of a simple
3D Fermi sea, the two lowest Landau bands below the Fermi level
are $(1)_{\downarrow}$ and $(0)_{\downarrow}$. Quantum
oscillations can still be observed at higher magnetic fiels (as
observed in transport), but no more variation of the polarization
due to the Landau quantization is expected.

\subsection{Non-zero temperature}

The temperature dependencies are calculated by including the Fermi
Dirac Distribution in the DOS integrations both for the simple 3D
DOS as well as for the LL modified DOS. The Fermi level then shows
the typical decrease at high temperature. If the temperature is
large enough compared to the Fermi energy, this leads at fixed
magnetic field to a drop of the polarization due to the
depopulation of the lower spin band, which can be seen in the
simulation in figure S\ref{Tdepshift} (dotted-line) which will be
discussed in the following section. If the sample Fermi energy is
very large, no effect on the spin polarization can be seen and the
Knight is relatively robust, as observed in the high density
sample P1-A1.

\newpage
\section{NMR properties}

\subsection{Methods}

The samples studied consist of thin flakes of Bi$_{2}$Se$_{3}$
crystal cleaved along the $(a,b)$ plane and their characteristics
are gathered in table \ref{tab1}. NMR measurements were initially
conducted simultaneously with transport (i.e. within the same
cooldown) on the central (contact-free) part of the sample to
establish uniform experimental conditions. The observed good
reproducibility of the sample properties from one cooldown to
another has later on enabled us to perform NMR before the
transport characterization to avoid potential perturbations due to
the electrical contacts, and optimize the filling factor of the
sample in the NMR coil. The magnetic field was calibrated by an
$^{27}$Al resonance of a $0.8 \mu$m thin aluminium foil placed
inside the sample coil. Its surface area was slightly greater than
the sample surface area (typically up to $3 \times 3$ mm) to
estimate the field inhomogeneity over the sample, which was found
to have a negligible effect on our results. In all our
experiments, even in our thinnest samples, the NMR signal
originates from the bulk of the sample due to the large ratio
between the bulk and surface nuclei. This was confirmed by the
absence of any thickness-dependant results in our study. The
spectra were collected by summing equally spaced FFT spectra
method~\cite{Clark1995} using the standard Hahn echo sequence,
$\pi/2 - \tau - \pi$. Because of the extremely thin coil thickness
(150 - 30 $\mu$m, to ensure the optimum filling factor) the
$\pi/2$ pulse width could be kept very small, about 1 $\mu s$. The
very short spin-spin relaxation time, $T_2$, and the oscillation
in the spin decay rate, \cite{Young2012} make the intensity
distribution of the satellites and the central line anomalous.
However, the anomaly is less severe if the spectra are recorded
with a very short delay time between the pulses.

\subsection{Angular dependence and extraction of isotropic and anisotropic shift for $^{209}$Bi NMR}

\subsubsection{Quadrupolar splitting}

$^{209}$Bi has a $\frac{9}{2}$ nuclear spin and thus shows four
pairs of satellite resonance lines along with the central line,
due to the quadrupolar coupling between the nuclei charge
distribution and the electric field gradients (EFG) in the
crystal. This can be seen in the NMR spectrum shown in figure 1 in
the main text, where the quadrupolar satellites are separated by
about 176 kHz. The position of the central line was determined by
a multiple Lorentzian fit of the entire spectrum with a fixed
quadrupolar splitting between the satellites. For P2-6, P3-6 and
P2-41 samples the quadrupolar splitting, f$_Q$, is 176$\pm2$ kHz,
while for P1 it is 160$\pm2$ kHz, comparable to the previously
reported values. \cite{Young2012,Nisson2013} We note that the
second order quadrupolar shift, which can lead to a shift of the
central line, is very small for such small f$_Q$ (typically $\sim
1$ kHz at the lowest Larmor frequency) and is neglected here. For
a quadrupolar nucleus in an electric field gradient, the satellite
lines are generally progressively broader than the central line
because the distribution of EFG due to defects leads to an
additional ``quadrupolar broadening''. This expected hierarchy in
the linewidth is not really observed in our samples, suggesting
that the broadening is rather dominated by some other intrinsic
factor. The observed global increase of the linewidth with the
magnetic field supports a magnetic origin which remains to be
clearly identified.

\subsubsection{Angular dependence of NMR}

\begin{figure}[!ht]
\includegraphics[width=15cm]{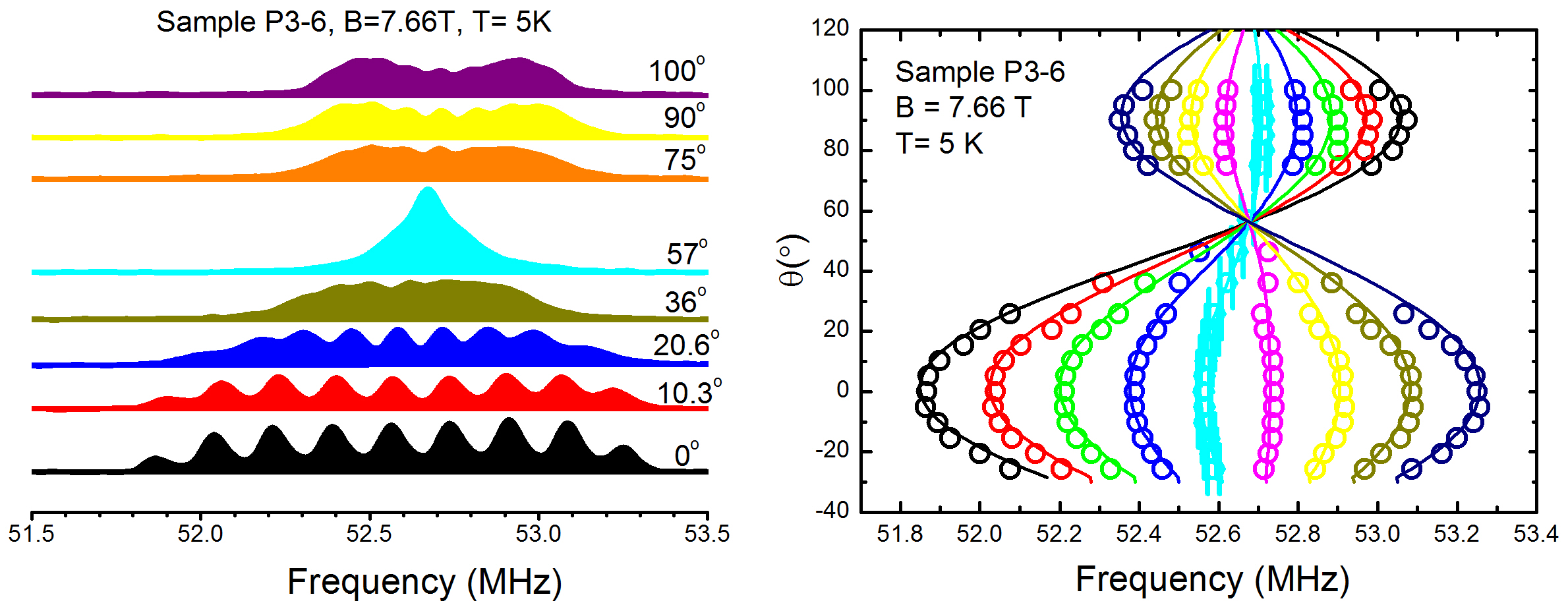}
\caption{Rotational pattern of the quadrupolar split $^{209}$Bi
spectra at $B=7.66$ T and $T=5$ K. (a) Evolution of the spectra
with the angle $\theta$. (b) Position of the resonances (open
circles) obtained by fitting each spectrum with 9 Lorentzian
functions, with proper constraints imposed by the spacing of the
satellites. The solid lines are a fit of the angular dependence of
the NMR spectrum with Eq.~\ref{anueqbis} (see
text).}\label{Bi2Se3NMRFigureNMRthetaSI}
\end{figure}

The angular dependence of the $^{209}$Bi NMR signal was studied in
sample P3-6 for magnetic fields up to 30 T. The results obtained
at $B=7.66$ T are reported in figure
S\ref{Bi2Se3NMRFigureNMRthetaSI}.
As the angle $\theta$ between the crystal \textit{c}-axis and the
magnetic field $B$ increases, the central line is right-shifted
(larger total shift), and the total Knight shift reaches a maximum
for $\theta=90^{\circ}$. The quadrupolar satellites follow a
particular angular dependence in which all the lines collapse into
the central line frequency for the ``magic angle''
$\theta=54.74^{\circ}$. To investigate this behavior more
quantitatively, we note that Bi$_{2}$Se$_{3}$ has an axially
symmetric EFG, with the principal axes of the EFG tensor
coinciding with the ones of the hyperfine shift tensor. The
angular dependence of the NMR line positions of a quadrupolar
split spectra can then be written as:

\begin{eqnarray}\label{angueq}
f(\theta)_{m\leftrightarrow(m-1)} &=& ^{209}\gamma B +
[\mathcal{A}_{iso}+(3\cos^2\theta-1)\mathcal{A}_{axial}] n_{e}
\times P(\theta)\\ \nonumber & & + \frac{f_Q} {2} \times
(m-\frac{1}{2})(3\cos^2 \theta -1)\\ \nonumber & & +
\frac{f_Q^2}{32\gamma B}(1-\cos^2\theta)\times \Big\{ \big [102 m
(m-1) -18 I (I+1) + 39 \big ]\cos^2\theta - \big [ 6 m (m-1) - 2 I
(1+1) +3 \big ] \Big \},
\end{eqnarray}

where $\mathcal{A}_{iso}$ and $\mathcal{A}_{axial}$ are the
isotropic and anisotropic hyperfine coupling constants,
respectively, $P(\theta)$ the electronic spin polarization,
$^{209}\gamma$ is the $^{209}$Bi gyromagnetic ratio, $I$ the
nuclear spin, $m$ the magnetic quantum number for $I$, and $f_Q$
is the nuclear quadrupole frequency. The first term (first line,
omitting the reference frequency $^{209}\gamma B$) is the same as
the one appearing in the main text and describes the frequency
shift of the central line. The axial term is weighed by the factor
$(3\cos^2 \theta -1)$ which originates from the integration on the
dipolar term in Eq. 1 in the main text. The second term (second
line) describes the angular evolution of the quadrupolar splitting
satellites. The last term in \{ .. \} (third line) is the second
order quadrupolar shift of the lines, which for our lowest Larmor
frequency is only $\sim 1 $ kHz and $\sim 6$ kHz for the central
line and fourth satellite transitions, respectively, and can
therefore be neglected. Eq. \ref{angueq} can be re-written as:

\begin{eqnarray}\label{anueqbis}
f(\theta)_{m\leftrightarrow(m-1)} &=& ^{209}\gamma B + \big [
\Delta f_{ \mathrm iso} + \frac{\Delta f_{\mathrm axial}}{2}
\times (3\cos^2\theta -1)\big] \times \frac{P(\theta)}{P(0)} \\
\nonumber & & + \frac{f_Q} {2} \times (m-\frac{1}{2})(3\cos^2
\theta -1),
\end{eqnarray}

where $\Delta f_{iso}=\mathcal{A}_{iso} n_{e} \times P(0)$ and
$\Delta f_{\mathrm axial}=2\mathcal{A}_{axial} n_{e} \times P(0)$
are the isotropic and axial component of the frequency shift at
$\theta=0^{\circ}$, respectively. The total frequency shift of
central line at $\theta=0^{\circ}$ is $\Delta f=\Delta
f_{iso}+\Delta f_{axial}$. Because of the anisotropy of the
conduction band $g$-factor (see I.C), the polarization appearing
in Eq. \ref{angueq} and \ref{anueqbis} may also be, depending on
the value of the magnetic field, anisotropic. This leads to a
$\theta$-dependant pre-factor which applies both to the isotropic
and axial component of the shift. We note that observing a central
line which is independent of $\theta$ in this system therefore
does not mean that the axial component is absent, but simply that
the $\theta$-dependence due to the axial term is compensated by
the anisotropic polarization. We have taken this effect into
account by including the $P(\theta)$ dependence calculated in our
model (see II. A) with $g^{*}_{eff}\perp = 23$, which we obtained
by using the $g$-factor anisotropy ratio observed in our low
density samples (see I.C), and $g^{*}_{eff} = 32.8$.

The fits of our experimental data using this equation, shown as
solid lines in figure S\ref{Bi2Se3NMRFigureNMRthetaSI}, are very
good. This enables us to determine the two fitting parameters,
$\Delta f_{ \mathrm iso}$ and $\Delta f_{\mathrm axial}$, for a
given magnetic field. In principal, these parameters are uniquely
defined by fitting the central line, since for $m=1/2$, the right
side of Eq. \ref{anueqbis} is reduced to the first term. Fitting
the satellites (whose angular dependence with respect to the main
line is known) further confirms the displacement of the central
line and enhances the fit accuracy. This process is repeated over
our total magnetic field range (6-30 T) \textit{to obtain the
magnetic field dependence of the isotropic and anisotropic
components of the Knight shift, which are reported in figure 3 in
the main text}. In high magnetic fields ($>$ 15 T) the rotational
pattern was only measured for 5-10 angles around
$\theta=0^{\circ}$, and at $\theta=54.74^{\circ}$ where all the
satellites collapse on to the central line, giving a better
precision for the fit. While the spectra were collected by
frequency sweeps in the superconducting magnet (6.1 - 17 T), most
of the high field spectra (15 - 30 T) were collected by field
sweep in the resistive magnet. For a field sweep spectrum, an
equation equivalent to Eq.~\ref{anueqbis} in the field domain was
used to find the isotropic and axial part of the shift.

\subsubsection{Isotropic shift}

As can be seen in figure 3 in the main text, the isotropic
component shows a clear saturation which can be perfectly
reproduced by our model. This enables us to extract the isotropic
hyperfine coupling constant $\mathcal{A}_{iso}$, and the density
of the electronic wave function at the nuclear site. We note that,
as mentioned before \cite{Willig1972,Miranda1974,Tunstall1988},
the determination of $\mathcal{A}_{iso}$ in the quantum limit is
highly reliable because the spin polarization is full and
independent of the band structure parameters. Our model further
shows that if the band parameters are well mastered (e.g. from
transport experiments), a single value of $\mathcal{A}_{iso}$ can
be used on the whole magnetic field range. Using our simple 3DEG
model (dotted line in figure 3 in the main text), one obtains
$\mathcal{A}_{iso}=7.41 \times 10^{-13} cm^{3}/s$, whereas the
more complete model including Landau quantization and disorder
(solid red line in figure 3 in the main text) gives
$\mathcal{A}_{iso}=6.94 \pm 0.16 \times 10^{-13} cm^{3}/s$, which
is the value we retain. The corresponding value $d_{\textrm{Bi}}$
of the density of the electronic wave function at the nuclear
site, is obtained as followed. From Eq. 1 in the main text, the
``contact'' term of the hyperfine coupling Hamiltonian is:

\begin{equation}\label{Hcontact}
\mathcal{H}_{contact}
=\frac{\mu_{0}}{4\pi}g_{0}\mu_{B}\gamma\hbar\textbf{I}\cdot
\frac{8\pi}{3}\textbf{S}\delta
(\textbf{r})=\frac{2}{3}\mu_{0}g_{0}\mu_{B}\gamma\hbar\textbf{I}\cdot
\textbf{S}\delta(\textbf{r})
\end{equation}

For a nucleus $i$ at a position \textbf{$r_{i}$}, this Hamiltonian
is equivalent to a Zeeman interaction in a magnetic field defined
by:
\begin{equation}\label{Bcontact}
\textbf{Be} =-\frac{2}{3}\mu_{0}g_{0}\mu_{B}\sum_{q}\textbf{S}_{q}
|\psi_{q} (\textbf{$r_{i}$})|^2,
\end{equation}

where the summation is done on the occupied electronic states of
wave function $\psi_{q}$. When the spin polarization of the
$n_{e}$ conduction electron is full, the corresponding frequency
shift is then:

\begin{equation}\label{freqshift}
\Delta f_{iso} =\frac{2}{3}\mu_{0}g_{0}\mu_{B}\gamma
(d_{\textrm{Bi}} \Omega) \frac{1}{2} n_{e},
\end{equation}

where $d_{\textrm{Bi}}$ is the density of the electronic wave
function \textit{at the nuclear site} normalized on the primitive
cell volume $\Omega$. The rhombohedral primitive cell in
Bi$_{2}$Se$_{3}$ has a volume $\Omega=1.497 \times10^{-23} cm^{3}$.
The factor $\frac{1}{2}$ originates from the projected value
${S}_{z}=\frac{1}{2}$ of the spin number in the fully polarized
regime. This equation and the definition of $\Delta f_{iso}$ in
III.B.2 defines the link between $\mathcal{A}_{iso}$ and
$d_{\textrm{Bi}}$, and we can deduce the value of
$d_{\textrm{Bi}}=8.72\times10^{25} cm^{-3}$.

We should point out that the extracted value of
$\mathcal{A}_{iso}$ and $d_{\textrm{Bi}}$ are  relevant for a
Fermi energy $E_{F}=19.7$ meV (sample P3-6), and may show a
dependence on the position of the Fermi level in the conduction
band. For the sake of simplicity, we have also omitted the
relativistic corrections which are known to enhance the value of
$\mathcal{A}_{iso}$ by a factor of the order of 3.
\cite{Sapoval1973} Additionally, the extracted value includes
possible ``core polarization'' effects. The core polarization
originates from an asymmetry of the spin part of the inner core
\textit{s}-shell wave function which develops due to the proximity
of unpaired electrons in the conduction band. For heavy elements,
this non-zero polarization can lead to a modification of the
``contact'' hyperfine coupling. In Bi$_{2}$Se$_{3}$ the unpaired
electrons are in the \textit{6p} shell with a possible
contribution from \textit{6s} electrons (this is discussed in the
main text). The core polarization shift arising from \textit{6p}
electrons is expected to be negative. \cite{carter1977} However,
contributions from \textit{s}-shells are always positive, and
generally considered as an amplification of the usual contact
term. The large positive isotropic shifts we observed in our work
suggest that core polarization effects, if important, would rather
arise from \textit{s} electrons. A better understanding of these
effects would require new theoretical studies and is out of the
scope of the present experimental paper. Finally, we would like to
discuss the possible role of spin-orbit effects in the isotropic
shift. It was theoretically proposed that spin-orbit effects could
lead to an additional contribution to the Knight
shift,\cite{Tripathi1982} which in some systems would manifest
itself as an enhancement of the isotropic shift. A signature of
this contribution is predicted to be visible in the low
temperature dependence of the Knight shift. In our samples, this
signature is absent, as shown in our study of the temperature
dependence of the Knight shift (see III.E), suggesting that this
spin-orbit contribution to the Knight shift is negligible in our
system. A last remark can be made about another effect of spin
orbit coupling, known as the ``$p_{1/2}$
corrections''.\cite{Larson2003} The $s_{1/2}$, $p_{1/2}$,
$p_{3/2}$, etc... orbitals are used to describe systems in the
presence of a strong spin-orbit coupling where the total momentum
J=L+S is the good quantum number. The $p_{1/2}$ state is formed by
admixing of an \textit{s}-type contribution to the \textit{p}
orbitals, and thus has a finite value at the nucleus site. In this
sense, the density of the electronic wave function at the nuclear
site $d_{Bi}$ discussed in our paper may not emanate from pure
\textit{s}-states only, but may also include the ``contact''
contribution of these $p_{1/2}$ orbitals.

\subsubsection{Anisotropic shift}

The anisotropic shift measured in our work essentially comes from
the dipolar interaction between the magnetic moments of the
$^{209}$Bi nuclei and the conduction electrons in the \textit{p}
orbitals. We note that the diamagnetic shift coming from the
orbital motion of the free electrons in a magnetic field is likely
to be negligible, since the ratio between this term and the
``contact'' term can be approximated in simple cases by
$\frac{1}{3}\frac{(m_{e}/m^{*})}{d_{\textrm{Bi}}\Omega}\sim
2\times 10^{-4}$.\cite{Hebborn1962}

The negative sign of the anisotropic term (see Fig.3 in the main
text), consistent with recent results \cite{Young2012}, suggests
that a significant fraction of the $p$ orbitals (more than
$\frac{2}{3}$) is of $p_{xy}$ rather than theoretically expected
$p_{z}$ nature. Recent results of Ref.~\onlinecite{Pertsova2013}
reveal the presence of a non-negligible $p_{xy}$ component (20 \%
at the CBM and 40 \% at the VBM) which could be at the origin of
our observation, even though this percentage is still too small to
account for a sign change in the dipolar term. We additionally
note that recent experiments performed on the Bi$_{2}$Se$_{3}$
surface states have also revealed a non-negligible ``in-plane''
component of the $p$ orbitals. \cite{Cao2013}

As can be seen in figure 3 in the main text, the magnetic field
behaviour of the anisotropic shift turns out to be more complex
than the one of the isotropic component. If only emanating from
the dipolar coupling, the anisotropic shift should also be
connected to the spin polarization (see Eq. \ref{anueqbis}) and
thus saturate above 16 T. An approximately linear increase with
almost zero intercept is indeed observed in the anisotropic shift
up to 15 T. However, the anisotropic shift keeps on increasing
above this field (leading to the drop clearly observed in the
total shift above 15 T in figure 1 of the main text) before
reaching an apparent saturation at around 22 T. At present we are
not able to give a satisfying explanation for this behavior. One
possibility is that, unlike the isotropic term, the dipolar term
exhibits a magnetic field dependence even when the spin
polarization is frozen because of magnetic-field-induced
modifications of the shape of the wave function.
This would however a priori not explain the saturating trend above
$B=22$T. Surprisingly, the field dependence of the anisotropic
shift can be very well reproduced using the equation $\Delta
f_{\mathrm axial}=2\mathcal{A}_{axial} n_{e} \times P$, where $P$
is calculated in our simple 3DEG model (see II.A) with a
$g$-factor \textit{$g^{*}_{eff} = 23$}, which is the value
characterizing the $g$-factor \textit{in the \textit{(a,b)}
plane}. A possible explanation for this ``delayed'' saturation,
could be the existence of a field dependence in the spin
polarization of the lowest Landau levels describing the
\textit{p}-wave states. This is precisely what is expected in the
general 3D Dirac Hamiltonian for Bi$_{2}$Se$_{3}$ where the
\textit{p}-wave orbitals are composed of a mixing of Bi and Se
\textit{p}-orbitals (see Ref.~\onlinecite{Orlita} for more
details). In this case, the second lowest Landau level tends to
depolarize with \textit{increasing} magnetic field, which reduces
the total spin polarization and delay the expected point (magnetic
field) for complete spin polarization. A deeper examination of
this high field behavior would require supplementary angular
dependent experiments and is out of the scope of the present
paper. The complex behavior of this anisotropic component further
illustrates the importance of being able to rely on the
\textit{isotropic} component to study the electronic spin
polarization.

\subsection{$^{77}$Se shift at low magnetic field}

The $^{77}$Se NMR, which is free from quadropular coupling effects
due to the $\frac{1}{2}$ nuclear spin, should \emph{a priori} have
been more suitable to investigate the shift of the main resonance
in Bi$_2$Se$_3$. However, a series of $^{77}$Se NMR measurements
has shown that: i) the relaxation times at low temperature are
very long (e.g. $\sim 0.5$ day at 60 K), which makes it more
complicated and time-consuming to study; ii) The Knight shift is
not as strong as it is for Bi, which was therefore better adapted
to probe the electronic properties of Bi$_2$Se$_3$.

\begin{figure}[!h]
\includegraphics[width=9cm]{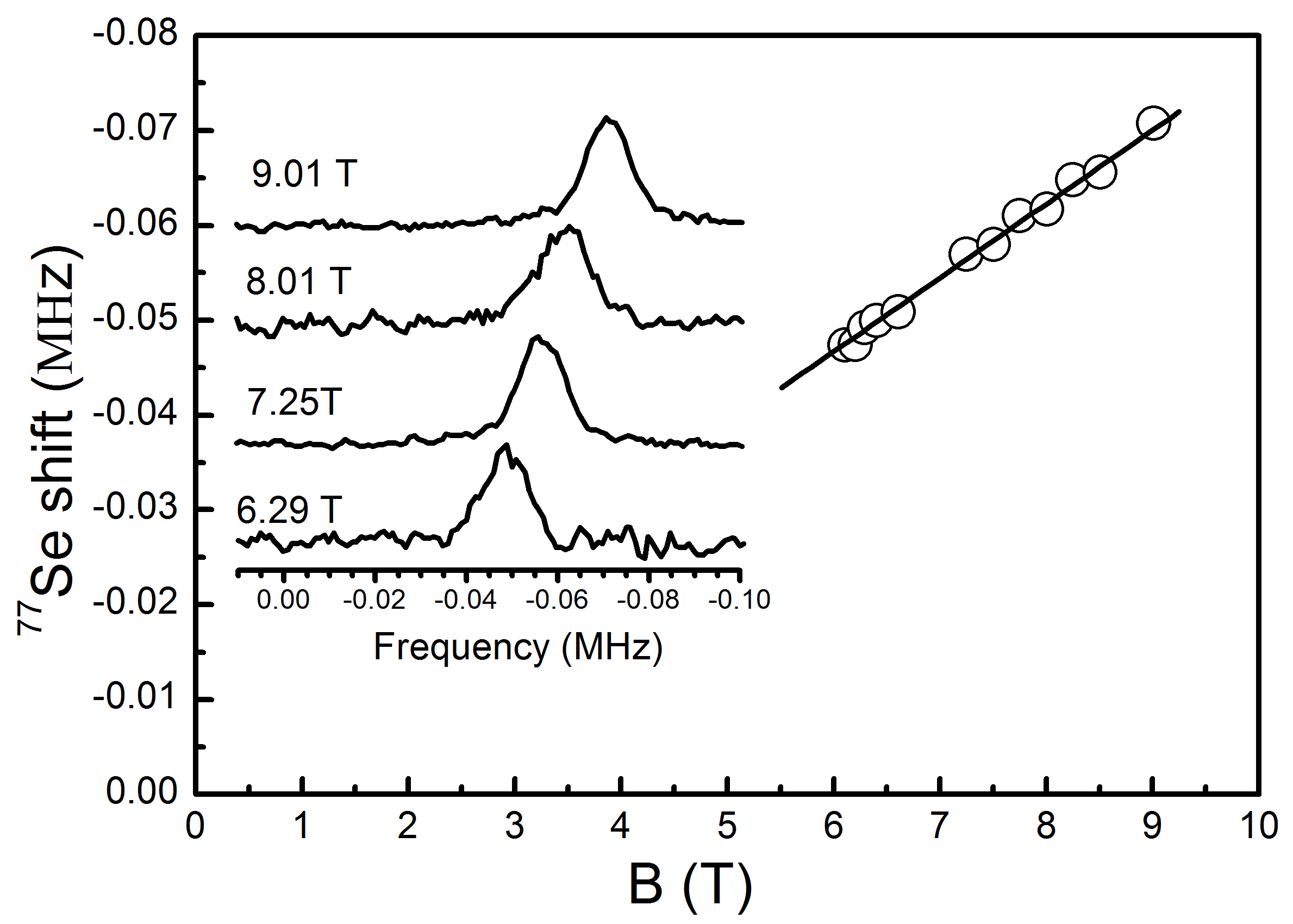}
\caption{\label{SeNMR} Field dependence of $^{77}$ Se shift at 1.5
K ($B \parallel$ \textit{c}-axis). The inset shows the
representative raw spectra at different magnetic fields.}
\end{figure}

The low field Knight shift at 1.5 K was nevertheless studied,
either by single shot, or using a small tipping angle pulse. In
spite of two inequivalent Se sites, we have observed only one Se
signal. Figure S\ref{SeNMR} shows the spectra and shifts of the
$^{77}$Se NMR up to 9 T, and at 1.5 K with $B \parallel $
\textit{c}-axis. The shift has been calculated with respect to the
bare Se gyromagnetic ratio, $^{77}\gamma=8.13$ MHz/T. \footnote{We
note that the $^{77}$ Se gyromagnetic ratio is not unambiguously
determined in the literature (see e.g.
Ref.~\onlinecite{Duddeck1995}), which can cause a significant
difference in the magnitude and even the sign of the resulting
shifts. This is at the origin of the large difference between the
shift reported in figure S\ref{SeNMR} and the one obtained in
Ref.~\onlinecite{Taylor2012}. The conclusion drawn from the
present value of the $^{77}$ Se shift is therefore based on the
assumption that the averaged value of 8.13 MHz/T is relevant in
our system.}

The low field frequency shift increases linearly, similarly to the
Bi shift, also pointing toward a connection to the electronic spin
polarization. The sign of the shift is however negative,
suggesting a ``dipolar'' rather than ``contact'' origin, which is
consistent with the expected absence of \textbf{s}-like wave
function at the Se site. \cite{Pertsova2013} This is also
responsible for the very long relaxation time observed. Further
experiments, including angular dependant NMR experiments, would be
required to fully characterize the $^{77}$Se NMR in Bi$_2$Se$_3$.
In particular, this could help our understanding of the behavior
of the dipolar coupling in high magnetic fields.

\subsection{density dependence of $^{209}$Bi NMR shift}

We report in this section our analysis of the density dependence
of the NMR shift at low magnetic fields.
\begin{figure}[!h]
\includegraphics[width=9cm]{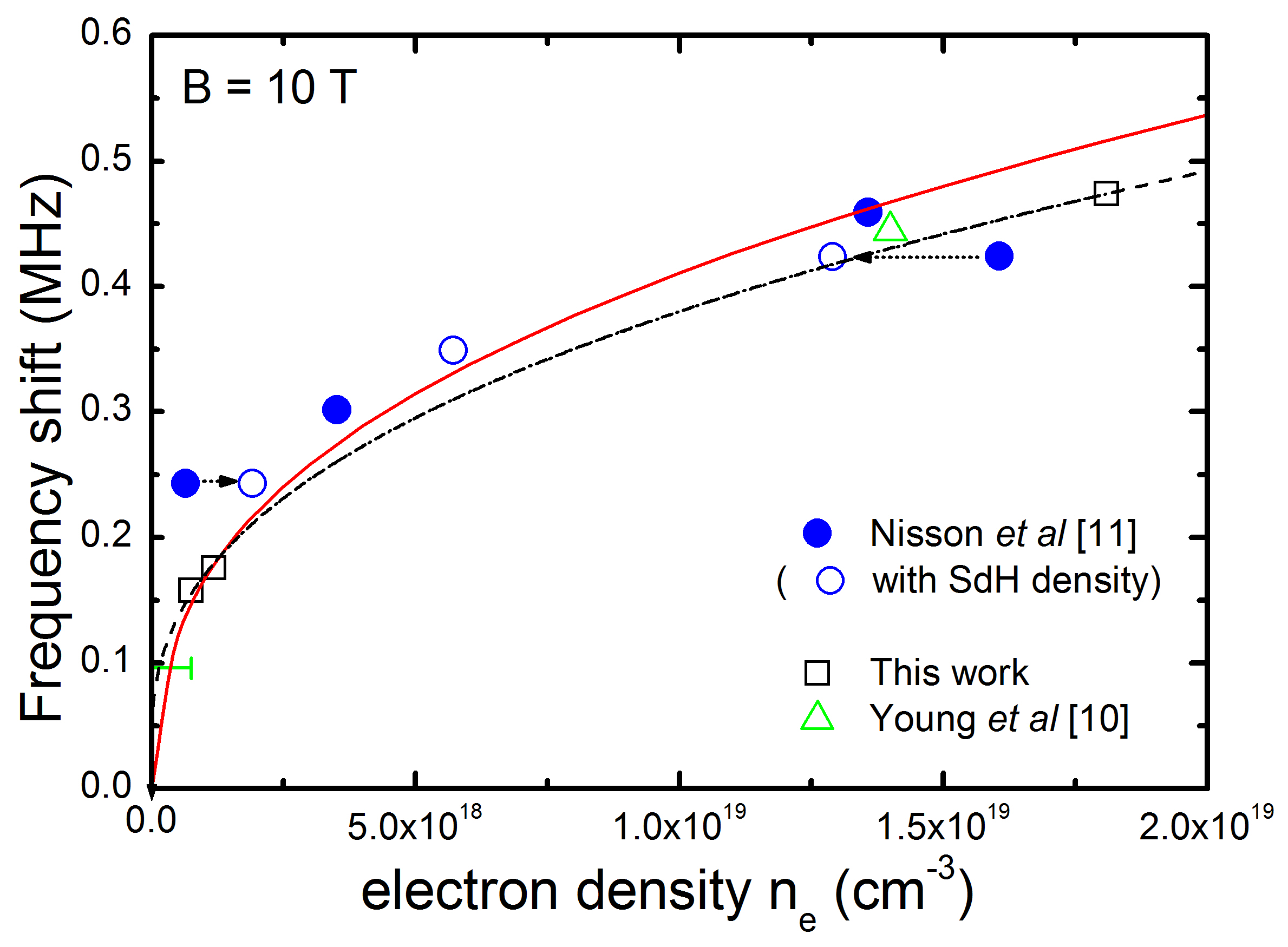}
\caption{Density dependence of the $^{209}$Bi central line shift
for our samples, together with results obtained in other recent
works. The horizontal dotted arrows show where the original data
points are located when the electron density is estimated from SdH
oscillations (see text). The solid line is the simulated variation
of shift in the 3DEG model discussed in II.A. The dashed line is a
fit to our data (black squares) allowing an extra non-zero
intercept.}\label{densdep}
\end{figure}
In figure S\ref{densdep}, we plot the total frequency shift at a
magnetic field of 10 T as a function of the electron density. This
magnetic field is low enough for all samples to be far from the
``saturating regime'' observed at high magnetic fields. We also
include in this plot data recently published in
Ref.~\onlinecite{Young2012,Nisson2013}. The shift is scaled to its
expected value at $B=10$ T assuming a linear field dependence in
this region, which is correct for electron densities higher than
the one of our sample P3-6 (for which the non-linearities of the
shift due to the proximity of the ``saturating regime'' will only
appear at higher magnetic fields). For the data of
Ref.~\onlinecite{Young2012}, the electron density was estimated
using the bulk Fermi energy determined from the ARPES measurement
reported therein. For their high density sample, we estimate
$E_{F}\sim 118$ meV, which using Eq. \ref{dens} gives $n_{e}\sim
1.4\times 10^{19} cm^{-3}$. For their low density sample, the bulk
conduction band is barely visible in the ARPES scan, which points
toward a very small electron density. We report the value of the
shift ($\sim 96$ kHz) as an horizontal bar of absciss coordinates
($n_{e}<7.47\times 10^{17}$) which corresponds to $E_{F}<18.5$
meV, a very large upper bound for the Fermi energy in this sample.
For the data of Ref.~\onlinecite{Nisson2013}, the electron
concentration was \textit{re-calculated} directly from the
oscillation frequency (as explained in I.B.) when available in the
table 1 of Ref.~\onlinecite{Nisson2013}. This generally gives a
more reliable estimation than the one obtained from the Hall
resistance which depends on various geometrical factors. The new
data points (shown as open circles in Fig. S\ref{densdep}) fall on
a common trend obeyed by all samples, showing a positive
correlation of the total NMR shift with the electron density. The
increase of the shift with the carrier concentration is close to
the previously reported $n_{e}^{1/3}$ behavior for $n$ and
$p$-type PbSe and PbTe. \cite{Leloup1973,SAPOVAL1968} In a
parabolic band approximation and at low magnetic fields, the
polarization of a 3D electron gas varies as $n_{e}^{-2/3}$, which
multiplied by the carrier concentration gives the $n_{e}^{1/3}$
dependence. Our simulation, plotted as a red solid line, shows the
exact quantitative result obtained when the anisotropy of the
conduction is taken into account. In this simulation (detailed in
II.A), the total NMR shift is directly connected to the spin
polarization with a single fitting prefactor $\mathcal{A}_{hyp}$
by writing $\Delta f=\mathcal{A}_{hyp}\times n_{e} \times P$.
$\mathcal{A}_{hyp}$ is an effective hyperfine coupling constant
taking into account the two main contributions (``contact'' and
``dipolar'') to the hyperfine coupling. The spin polarization has
been calculated with $g^{*}_{eff} = 32.8$ obtained for sample P3-6
by the field dependence of the NMR shift (see main text). Given
the fact that each sample may present a different amount of
disorder and a possible density-induced variation of the
$g$-factor, this model catches the experimental data surprisingly
well. This agreement not only shows that the NMR shift is
essentially coming from the coupling to the conduction electrons,
but also suggest that their degree of polarization is directly
present in the total shift, as expected from the nature of the
hyperfine coupling (see Eq. \ref{anueqbis}). The extracted value
of the parameter $\mathcal{A}_{hyp}$ which best describes all the
existing data is $\mathcal{A}_{hyp}=2.76 \pm 0.25 \times 10^{-13}
cm^{3}/s$, where the error bar is estimated from the scattering of
the experimental data around the simulated curve. For a more
complete modelling of the total shift, we have also tried to fit
our experimental data allowing an extra non-zero intercept in
addition to our calculated hyperfine shift, which represents the
effect of a possible additional ``chemical shift'', independent of
the conduction electrons. The fit of our data in these conditions
(dashed line) yields $\mathcal{A}_{hyp}=2.34 \times 10^{-13}
cm^{3}/s$ and an intercept (residual shift) of 33 kHz, which is
still an order of magnitude below the isotropic hyperfine
component of main interest in this work. This very small intercept
in the zero-density limit rules out the presence of isolated
magnetic moments in the studied samples, as well as a significant
direct contribution of filled electronic shells (chemical shift).

We finally note that the simplicity of this observed dependence on
density, which can be very useful to characterize the
\textit{total} NMR shift in Bi$_2$Se$_3$, can to some extent be
surprising. Indeed, according to the first term of Eq.
\ref{anueqbis} taken at $\theta =0^{\circ}$, the density
dependence of the total frequency shift is given by the factor
$n_{e} \times P$ provided
($\mathcal{A}_{iso}+2\mathcal{A}_{axial}$) is independent of the
electron density. On the other hand, the density dependence of the
isotropic and anisotropic shifts, taken separately, differs from
the $n_{e}^{1/3}$-like behaviour expected if each coupling
constant $\mathcal{A}_{iso}$ and $\mathcal{A}_{axial}$ was
independent of the density. As a simple example, the anisotropic
shift measured for P1-3 (high carrier density) is much smaller
than its value at lower electron density, which implies that
$\mathcal{A}_{axial}$ is not constant but decreases with $n_{e}$.
More generally, the ratio between the isotropic and anisotropic
component of the Knight shift increases with the Fermi energy (or
carrier concentration), in qualitative agreement with
Ref.~\onlinecite{Young2012}. This points toward a change in the
nature of the wave function as we go to higher energies in the
conduction band, characterized by an increase of the \textit{s}
character. Our observations in figure S\ref{densdep} suggest that
the density dependence of $\mathcal{A}_{axial}$ and
$\mathcal{A}_{iso}$ are apparently compensated, such that
$\mathcal{A}_{iso}+2\mathcal{A}_{axial}$ has a weak density
dependence. Whether this is a simple coincidence or an effect
rooted to some deeper physical arguments is an open question which
is left for future works.

\subsection{Temperature dependence of $^{209}$Bi central line shift}

\begin{figure}[!h]
\includegraphics[width=9cm]{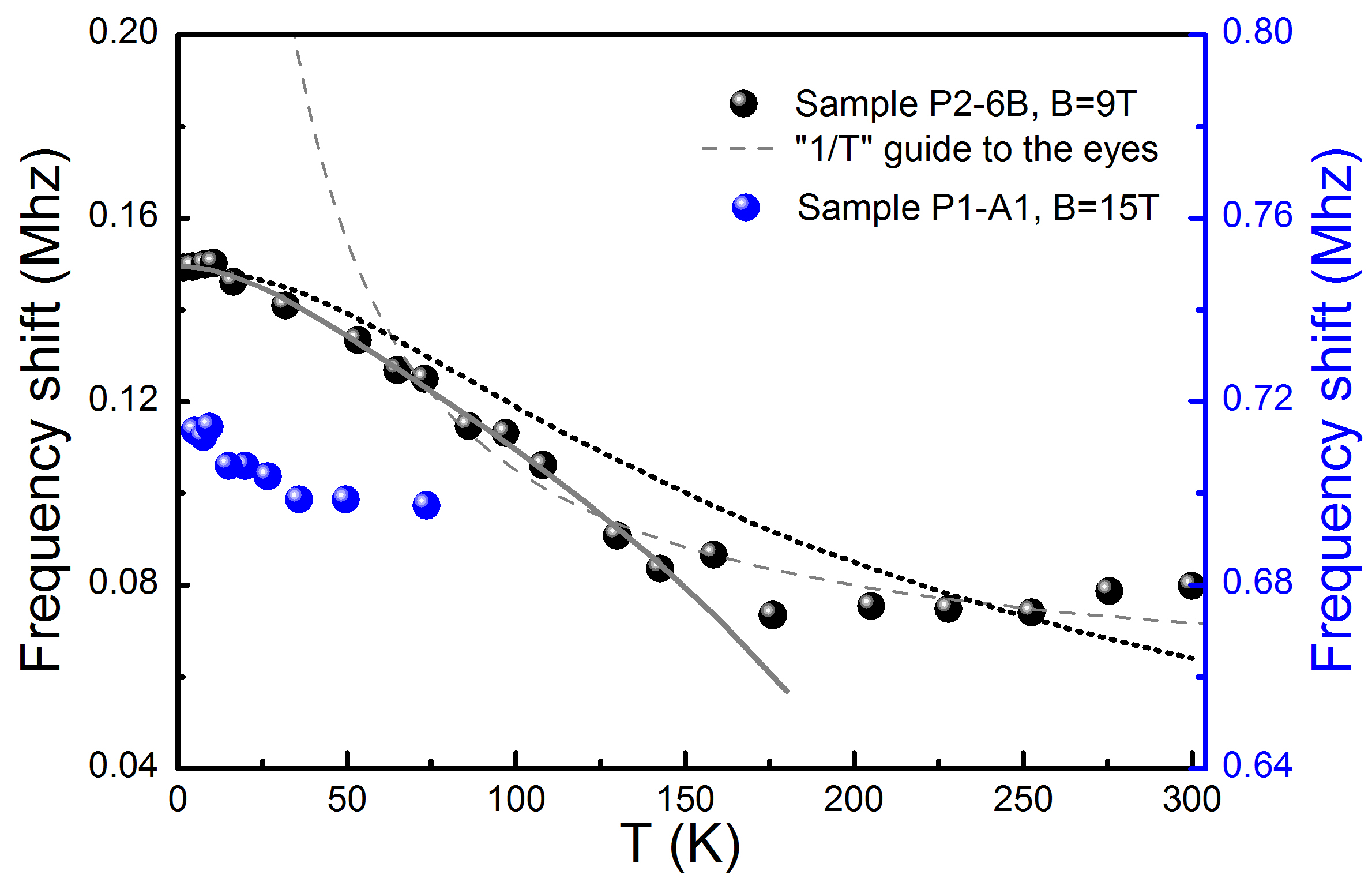}
\caption{\label{Tdepshift}Temperature dependence of the $^{209}$Bi
frequency shift for samples P2-6B and P1-A1 at $B=9$ and 15 T,
respectively. The dotted-line is the simulated variation of the
shift using our 3DEG model (see text). The solid line further
includes the effect of disorder. The dashed-line is a guide to the
eyes for a $1/T$ behavior.}
\end{figure}

We have performed temperature dependent NMR measurements on
different samples (P2-6B,P3-6,P1-A1) for magnetic fields up to 30
T. In figure S\ref{Tdepshift}, we report the variation of the NMR
central line frequency shift as a function of temperature up to
room temperature at magnetic fields of $B=$9 T and $B=$15 T for
sample P2-6B, and P1-A1, respectively. In both cases, these
magnetic fields are well below the magnetic field $B_{s}$
corresponding to the full electronic spin polarization. The P2-6B
shift (left-scale) is constant a low temperature (2 K-30 K), and
then smoothly drops until $T\sim 200$ K where it tends to
saturate. Sample P1-3 (right scale) shows a similar but less
pronounced trend.

We note that a shift emerging from localized magnetic moments
should exhibit a characteristic $1/T$ Curie temperature dependence
(shown as a dashed-line guide to the eyes in figure
S\ref{Tdepshift}) which is clearly not observed in any of our
samples at low temperature. On the other hand, the simulation of
the temperature dependence of the Knight shift obtained with our
3DEG model (dotted-line, see section II.C. for  the calculation
details) gives a good qualitative description of the shift
evolution for temperatures up to $\sim 100$ K. This shows that the
temperature dependence of the total frequency shift can be
attributed to the temperature dependence of the spin polarization,
essentially coming from the smearing of the Fermi-Dirac function
at the Fermi level. Including disorder (solid line), and in
particular its increase with temperature inferred from the
zero-field resistance, gives a very good description of the data,
at least up to $\sim$150 K. We note that a possible spin-orbit
induced contribution to the shift \cite{Tripathi1982} should lead
to a different temperature dependence \cite{Misra1987} which is
not observed in our experiment. It is also not possible to fit our
shift correctly with a temperature-independent term, which rules
out the presence of a significant ``spin-independent'' orbital
coupling, at least at these magnetic fields.

At higher temperature, the shift stops dropping and even shows a
slight increase above 200 K. This behavior can be attributed to
the carrier activation through the Bi$_{2}$Se$_{3}$ energy gap
($0.215\pm5$ meV, optically measured on the same samples), which
can independently be observed with transport (Hall) measurements
at high temperatures. This slight increase of the shift is
therefore another manifestation of the dependence of the shift on
the total carrier density. The smoother variation of the shift
observed for sample P1-A1 is expected, since the higher Fermi
energy in this sample makes the polarization more robust with
respect to temperature effects. The high magnetic field
temperature-dependence of the total shift has also been measured
and shows a more complex behavior due to the combination of the
different behavior of the isotropic and axial components in high
fields.